\documentclass[11pt,fleqn]{article}

\usepackage{graphicx}
\usepackage[a4paper,left=1in,right=0.5in,top=0.8in,bottom=0.8in]{geometry}
\usepackage{amsmath}
\usepackage[authoryear]{natbib}
\usepackage{txfonts}
\usepackage{subcaption}
\usepackage{lscape}
\usepackage{placeins}
\usepackage{xcolor}
\usepackage{bm}
\usepackage{url}

\newcommand{\solrat}{\textsc{SolRaT}}
\newcommand{\tablefootmark}[1]{\textsuperscript{#1}}
\newcommand{\tablefoottext}[2]{\textsuperscript{#1}#2\par}
\newcommand{\tablefoot}[1]{\par\smallskip{\footnotesize #1}}

\begin{document}

\title{SolRaT: polarized spectral line modeling with multi-term
and multi-level atoms}

\author{
I. I. Yakovkin\\
\\
\small{Institute of Physics of the National Academy of Sciences of Ukraine, Kyiv, Ukraine}\\
\small{Astronomical Observatory of the Taras Shevchenko National University of Kyiv, Kyiv, Ukraine}
}
\date{}

\maketitle

\begin{abstract}
\noindent\textbf{Context.}
The interpretation of polarized spectral lines requires models that
    combine statistical equilibrium with polarized radiative
    transfer accounting for atomic polarization, the Hanle effect, and
    Zeeman and Paschen-Back splitting.

\noindent\textbf{Aims.}
The aim is to provide an open-source Python implementation of the
    irreducible spherical tensor formalism of
    \citet{landi2004polarization}, hereafter LL04, for forward synthesis of Stokes
    profiles with interchangeable atomic descriptions.

\noindent\textbf{Methods.}
\solrat\ solves the statistical equilibrium and polarized transfer
    problem for interchangeable multi-term and multi-level atomic
    descriptions. The multi-term model retains coherences between
    fine-structure levels of the same term and treats the magnetic
    Hamiltonian by exact diagonalization, allowing the magnetic splitting
    to be followed from the linear Zeeman regime to the incomplete and
    complete Paschen-Back regimes. The widely used multi-level model in turn assumes only the linear Zeeman regime,
    resulting in a simpler and more developed formalism.

\noindent\textbf{Results.}
The implementation is checked against analytic results from LL04,
    the Unno-Rachkovsky solution for local thermodynamic equilibrium
    (LTE) Zeeman transfer, the
    \textsc{Hazel2} He\,\textsc{i}~D$_3$ synthesis, and published
    self-consistent non-LTE scattering benchmarks. The examples show how
    the intra-term magnetic mixing produces departures from a
    linear Zeeman multi-level description, and how a
    $J$-constrained multi-term approach can be used to isolate this effect on a
    chosen fine-structure component.

\noindent\textbf{Conclusions.}
Synthesizing the same spectral line with interchangeable multi-term and
    multi-level descriptions under identical conditions isolates the signatures of
    intra-term magnetic mixing and inter-$J$ coherences from the linear
    Zeeman response. Such controlled comparisons are central to developing
    scattering polarization diagnostics applicable across multiple
    atomic descriptions.
\end{abstract}

\vspace{2em}

\section{Introduction}

Spectropolarimetry is widely used to probe the vector magnetic field in the solar atmosphere through the dependence of
the Stokes parameters $I$, $Q$, $U$, and $V$ on field strength, geometry, and the local thermodynamic state.
The magnetic field leaves its imprint on the polarization of spectral lines
through the splitting of the atomic sublevels and through the way atoms scatter the
incident radiation. Measuring all four Stokes parameters therefore allows to infer the
field strength and orientation. The
photosphere, the chromosphere, and the corona each contribute to different spectral lines
under different magnetic regimes.
This makes polarimetry the primary remote diagnostic for magnetic field in solar atmosphere.

Interpreting the Stokes profiles requires the simultaneous solution of the polarized radiative transfer equations (RTE)
and the statistical equilibrium equations (SEE) for
the atomic populations \citep{de2017radiative}.
Departures from local thermodynamic equilibrium (LTE), hereafter non-LTE effects, alter both the atomic populations and
the atomic level polarization
and are essential for interpreting observed spectra \citep{shchukina1997nlte}.
The populations of the atomic levels depend on the radiation field, and the
radiation field depends in turn on the populations through the emission and absorption
along the ray. Both parts have to be solved together once the line becomes optically
thick. The polarization adds further components to the same system, because the
sublevels are populated at unequal rates and the transfer couples the Stokes parameters
to each other.

In regimes where atomic level coherences are significant, the density matrix is often written in the irreducible
spherical tensor formulation.
The evolution of the resulting statistical tensors is dictated by the radiation field and by the magnetic Hamiltonian.
The corresponding formalism was developed by \citet{landi2004polarization} (hereafter LL04)
and is adopted throughout this paper.
Related Stokes-vector formulations for polarized transfer in magnetized media are discussed, for example,
in \cite{judge1998spectral}.

The magnetic regimes support different diagnostics.
The linear Zeeman effect underlies photospheric magnetometry: for splitting small compared to the Doppler width, 
the weak-field approximation relates Stokes $V$ to the longitudinal field component
and to the wavelength derivative of Stokes $I$,
and for resolved splitting the Stokes vector constrains the field vector through inversion \citep{centeno2018weak}.
Once the magnetic splitting becomes comparable to the fine-structure separation within a term,
the atom enters the incomplete Paschen-Back regime,
in which the component frequencies and strengths become field dependent,
and a linear Zeeman analysis of a single component loses accuracy \citep{socas2004signatures}.
The Hanle effect covers a complementary range, responding to fields that leave a weak Zeeman signature,
including tangled fields for which the circular polarization cancels within
the resolution element \citep{stenflo1982hanle, bueno2004substantial, bueno2009three}.
It is widely applied to corona and chromosphere \citep{trujillo2017physics, stenflo1997second, gandorfer2005second}.
In the corona the forbidden magnetic-dipole lines can form in the saturated Hanle regime, where the linear polarization 
is sensitive to the field direction \citep{casini1999spectral},
which has been used to measure the coronal field strength directly \citep{lin2000new}.
The density matrix formulation adopted here treats the atomic populations, the polarization, and the coherences 
within the same statistical equilibrium system, taking all mentioned effects into account.

Recovering the field vector from an observed Stokes profile requires a
forward model that reproduces the profile from a specified atmosphere. Closed-form
relations such as the weak-field approximation and the Unno-Rachkovsky solution for the Milne-Eddington atmosphere serve
this role while the splitting stays linear and the populations stay close to thermal.
Once atomic polarization, inter-$J$ coherences, or field-dependent component strengths
contribute, the coupled equilibrium and transfer problem can be solved only numerically. 
A number of numerical codes were developed to help interpreting the observed spectra
under different conditions and approximations.
\textsc{Hazel} and \textsc{Hazel2} \citep{asensio2008advanced} provide forward modeling
and inversions for selected chromospheric slab diagnostics, in particular the helium multiplets.
\textsc{Porta}~\citep{vstvepan2013porta} solves scattering polarization self-consistently
in three-dimensional model atmospheres with a parallel multi-ray formal solver
and targets realistic three-dimensional synthesis.
\textsc{P-Corona}~\citep{hebbur2025p} applies the LL04 density matrix formalism to the intensity
and polarization of forbidden and permitted coronal lines in three-dimensional models of the solar corona,
accounting for anisotropic radiation pumping and the Hanle and Zeeman effects.
\textsc{RH}~\citep{uitenbroek2001multilevel} addresses multi-level non-LTE transfer
with partial frequency redistribution
(PRD) \citep[][and following papers in that series]{bommier1997master, bommier1997master2} and Zeeman polarization.
Codes such as \textsc{SIR}~\citep{ruiz1992inversion}, \textsc{SPINOR}~\citep{frutiger2000inversions},
\textsc{Nicole}~\citep{socas2015open},
and \textsc{STiC}~\citep{de2019stic} are designed to automate the inference of atmospheric stratifications
from observations by performing a robust
 inversion of Stokes profiles \citep[for a more detailed review see][]{del2016inversion}.

Existing codes for polarized spectral line modeling are each optimized for a different scientific task.
Each code fixes a particular atomic treatment, a particular transfer scheme,
and a particular set of approximations, chosen for performance in its intended
application. A difference between two published syntheses of the same line therefore
simultaneously mixes all differences in the physical assumptions that distinguish the two codes. 
Isolating the effect of one assumption requires holding the atmosphere, the
incident radiation, the geometry, and the transfer solver fixed while that one
assumption is varied. Separating the impact of individual physical
assumptions therefore requires a single code that allows for a
gradual migration between multiple radiative transfer problems.

\solrat\ is designed as an inspectable implementation of the LL04 density matrix formalism for forward synthesis.
The multi-term and multi-level descriptions are implemented within the same radiative transfer formulation.
The multi-term atom follows the angular momentum algebra of LL04 and diagonalizes the magnetic Hamiltonian
within each term, while the multi-level atom uses empirical level energies and Land\'e factors.
Applying these descriptions to the same atmosphere, incident radiation field, magnetic geometry,
and line of sight makes it possible to identify the emergent-profile features that require inter-$J$ coherences,
Paschen-Back mixing, or empirical line data.
Such controlled comparisons are essential for developing new polarized-line
diagnostics sensitive to selected physical mechanisms.

\section{Two atomic descriptions}

\solrat\ solves the RTE and the SEE of LL04 and follows its notation throughout.
The polarization state of the radiation is described by the four Stokes parameters
collected in the column vector $\mathbf I=(I,Q,U,V)^T$.
The atomic state is described by the statistical tensors $\rho^K_Q$ of the density matrix,
where $K$ is the tensor rank and $Q\in\{-K,\dots,+K\}$ is the tensor component.
The statistical tensors collect the level populations together with the coherences between magnetic sublevels.
\solrat\ exposes two complementary atomic descriptions, multi-term and multi-level,
through the same abstract interface.
The two descriptions differ in how the atomic state and the rates coupling it to the radiation field are built.
Both keep the coherences between magnetic sublevels and therefore reproduce the Hanle effect,
while the multi-term description also retains coherences between the fine-structure levels of a term,
which can affect the scattering polarization of chromospheric resonance multiplets \citep{del2020magnetic}.

In the multi-term description, a term is identified by the quantum numbers $(\beta, L, S)$,
where $\beta$ labels the inner electronic configuration and $L$, $S$ are the orbital and spin angular
momenta, respectively.
Within a term, levels are indexed by $J\in\{|L-S|,\dots,L+S\}$ with magnetic sublevels $M\in\{-J,\dots,+J\}$.
Coherences $\rho^K_Q(J,J')$ between fine-structure levels $J$ and $J'$ of the same term are retained.
The corresponding SEE and RTE are described in Appendix~\ref{sec:app_equations}.

The energies of the magnetic sublevels of a term are obtained by diagonalizing the spin-orbit and magnetic Hamiltonians
together within that term, resulting in field-induced mixing of its fine-structure levels.
For a magnetic field $\bm B$ of strength $B$ along the quantization axis, the total Hamiltonian takes the form
\begin{equation}
  \hat H = \hat H_0 + \hat H_{SO} + \hat H_B\,,
  \label{eq:H_full}
\end{equation}
where $\hat H_0$ is the unperturbed Hamiltonian of the term, $\hat H_{SO}$ is the spin-orbit Hamiltonian,
and $\hat H_B$ is the magnetic Hamiltonian.
The spin-orbit Hamiltonian can be written as
\begin{equation}
  \hat H_{SO} = \zeta(\beta L S)\,\hat{\bm L}\cdot\hat{\bm S}\,,
  \label{eq:H_SO}
\end{equation}
where $\zeta(\beta L S)$ is the spin-orbit coupling constant of the term,
and $\hat{\bm L}$ and $\hat{\bm S}$ are the orbital and spin angular momentum operators.
The magnetic Hamiltonian is
\begin{equation}
  \hat H_B = \mu_B B\,(\hat J_z + \hat S_z)\,,
  \label{eq:H_B}
\end{equation}
where $\mu_B$ is the Bohr magneton, and $\hat J_z$ and $\hat S_z$ are the components
of the total and spin angular momentum operators along the quantization axis.
At fixed $M$, the matrix of $\hat H_{SO}+\hat H_B$ is tridiagonal in $J$.
Its diagonal elements give the fine-structure energies and the linear Zeeman shifts,
and its off-diagonal elements couple neighboring $J$ values.
Exact diagonalization gives the field-dependent Paschen-Back eigenvalues $\tilde E_{j,M}$ and eigenvectors $c_{j J M}$.
Here $j$ labels the eigenstates at fixed $M$, and $c_{j J M}$ is the amplitude of the level $J$ in the eigenstate $j$.
The matrix elements used by \solrat\ are given in Appendix~\ref{sec:app_hamiltonian}.

\begin{figure}
  \centering
  \includegraphics[width=0.5\linewidth]{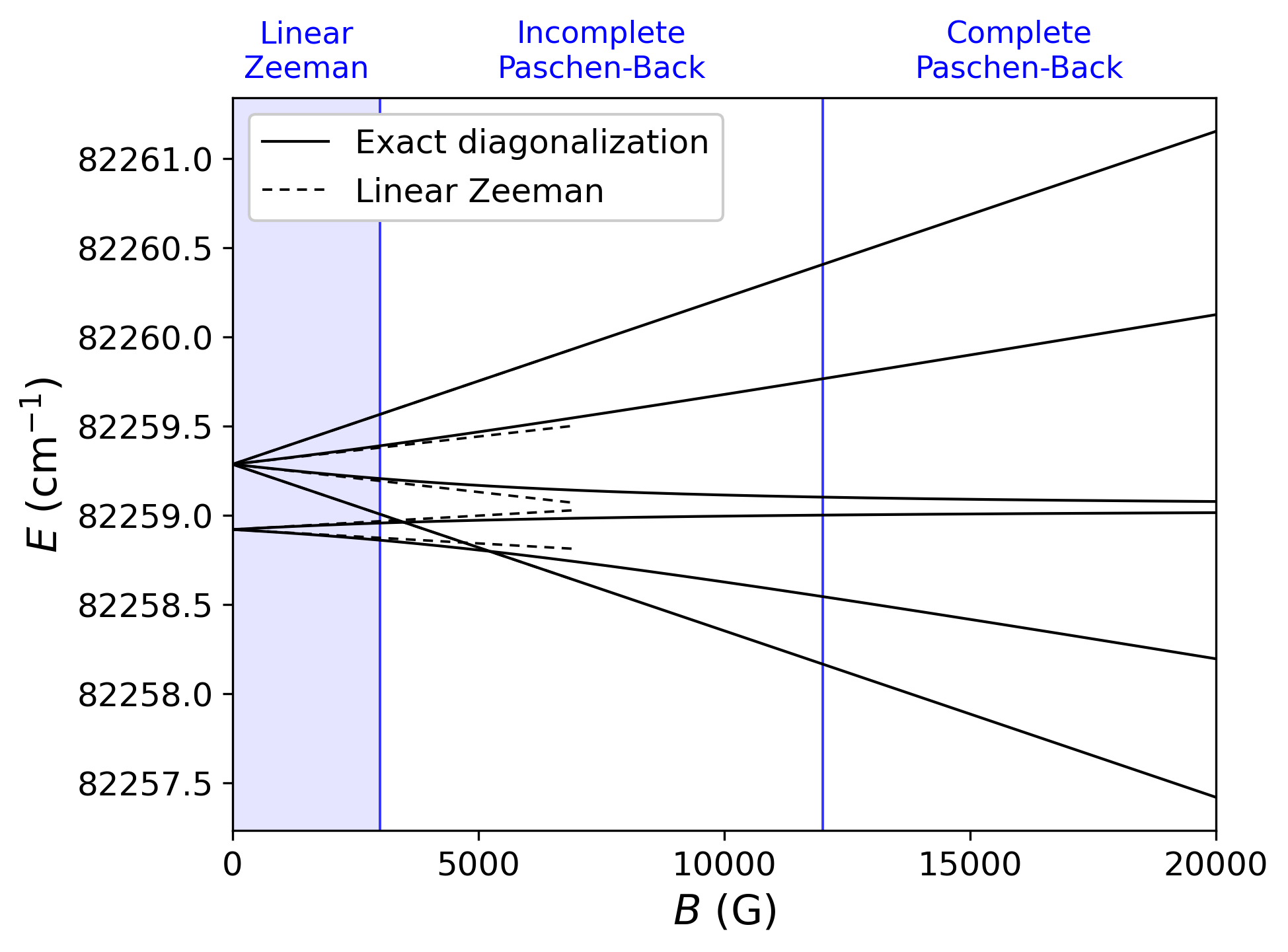}
  \caption{Magnetic sublevel energies of the hydrogen $2p\,^2P$ term from
    0 to 20~kG. Solid lines: exact diagonalization. Dashed lines: linear Zeeman
    energies, drawn over the lower part of the range. Vertical lines mark
    the approximate regime boundaries, and the linear Zeeman range is shaded.}
  \label{fig:paschen_back}
\end{figure}

For weak fields the off-diagonal couplings are negligible and the spectrum reduces to the linear Zeeman pattern.
With increasing $B$, neighboring $J$ values mix,
and the component frequencies and oscillator strengths become field dependent.
The relative strengths of the $\pi$ and $\sigma^\pm$ components then depart from the linear Zeeman pattern,
and the atom enters the incomplete Paschen-Back regime.
In the high-field limit the atom approaches the complete Paschen-Back regime.
For an order-of-magnitude estimate, the boundary between the linear Zeeman and incomplete Paschen-Back regimes
is $B_{\rm PB}=\Delta E_{\rm FS}/(g_J\mu_B)$, where $\Delta E_{\rm FS}$ is the smallest fine-structure separation
participating in the observed component and $g_J$ is the corresponding weak-field Land\'e factor.
Figure~\ref{fig:paschen_back} shows the magnetic sublevel energies of the hydrogen $2p\,^2P$ term,
the lower term of H$\alpha$, with its two fine-structure levels separated by approximately $0.37$~cm$^{-1}$.
The solid curves are the eigenvalues of the diagonalized Hamiltonian
and the dashed curves are the linear Zeeman energies.
The solid and dashed curves agree within the shaded weak-field range,
where the linear Zeeman treatment is adequate, and separate once the magnetic splitting becomes
comparable to the fine-structure separation.

In the multi-level description, a level is identified by a label $\alpha$ and by $J$,
with $\alpha$ collecting the remaining quantum numbers.
Coherences between levels of different $\alpha$ or $J$ are omitted,
and the statistical tensor of each level reduces to $\rho^K_Q(\alpha,J)$.
The corresponding SEE and RTE are described in Appendix~\ref{sec:app_equations}.
Each level carries an empirical energy $E_{\alpha J}$ and Land\'e factor $g_{\alpha J}$,
both supplied by the user.
The magnetic splitting is the linear Zeeman shift $g_{\alpha J}\mu_B B M$.
A multi-term atom reduces to this form when the inter-$J$ coherences are omitted and
the magnetic structure is kept in fixed-$J$ linear Zeeman levels.
This description is appropriate when the magnetic splitting stays well inside the linear Zeeman range
(Fig.~\ref{fig:paschen_back}), or when a study targets the thermodynamic and velocity structure
of the atmosphere instead of its magnetic fields.
The multi-level description is also simpler, because it carries one statistical tensor per level
and needs only an energy and a Land\'e factor for each level.
Codes implementing the multi-level description are therefore generally faster, and the multi-level
methodology is more developed, for example in non-LTE transfer with PRD \citep{uitenbroek2001multilevel}.

\begin{figure}
  \centering
  \includegraphics[width=0.5\linewidth]{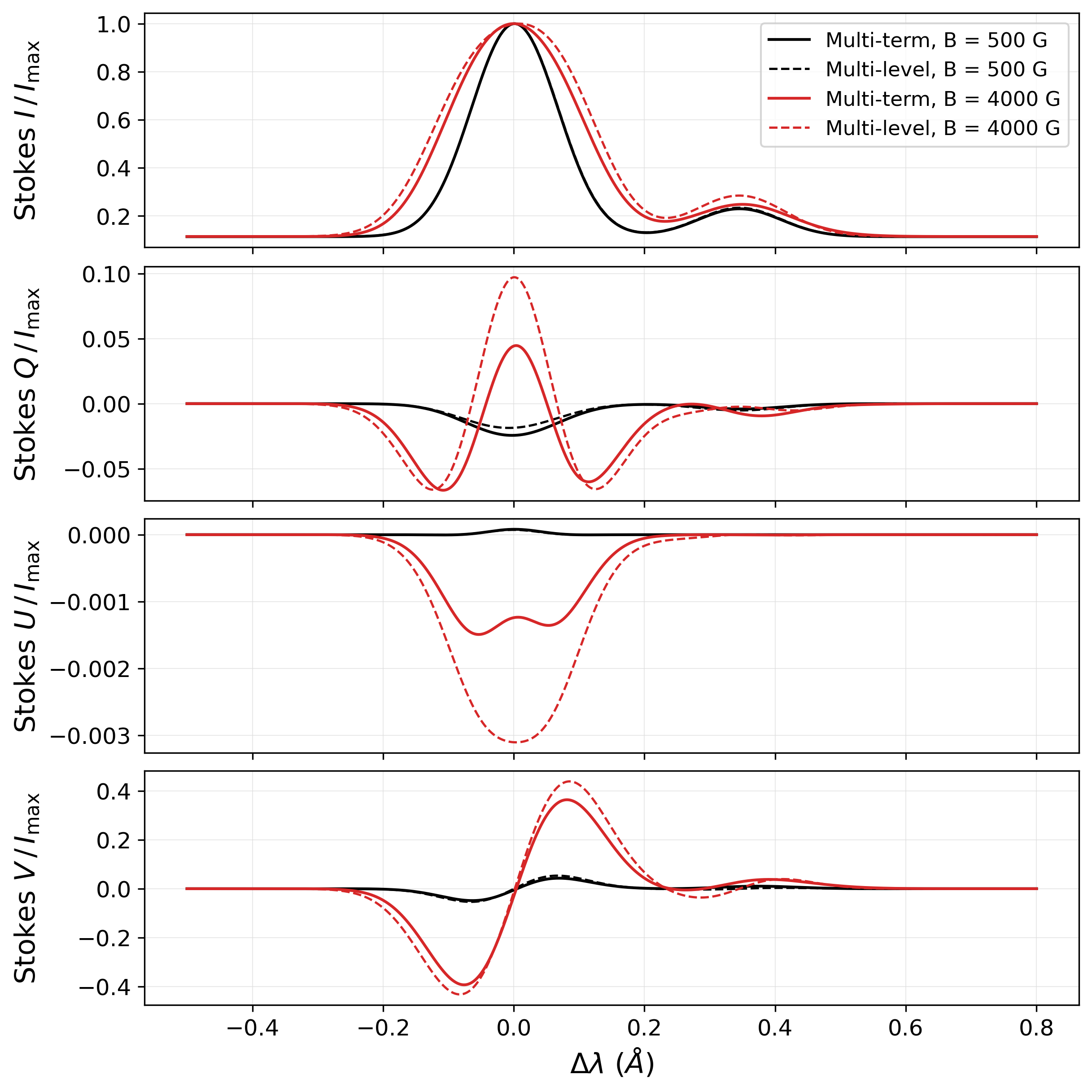}
  \caption{He\,\textsc{i}~D$_3$ Stokes profiles normalized by
    $I_{\max}$ for $B=500$~G and $4$~kG. Solid lines: multi-term.
    Dashed lines: direct multi-level analog.}
  \label{fig:HeI_Stokes}
\end{figure}

Figure~\ref{fig:HeI_Stokes} shows He\,\textsc{i}~D$_3$ Stokes profiles computed with the multi-term atom and
with its direct multi-level analog in a constant-property slab illuminated by a prescribed radiation field.
At $B=500$~G, Stokes $I$ and $V$ remain close between the two descriptions, while the linear-polarization
profiles can already show visible differences.
As the field magnitude increases to $B=4$~kG, the incomplete Paschen-Back mixing in the multi-term atom produces
a substantially different Stokes vector: the multi-level profile is broader in Stokes $I$ and reaches larger
peak amplitudes in both linear and circular polarizations.
The transition is continuous because the magnetic Hamiltonian of each term is diagonalized at every field value.

\section{Forward-synthesis workflow}
\label{sec:workflow}

\begin{figure}
  \centering
  \includegraphics[width=\linewidth]{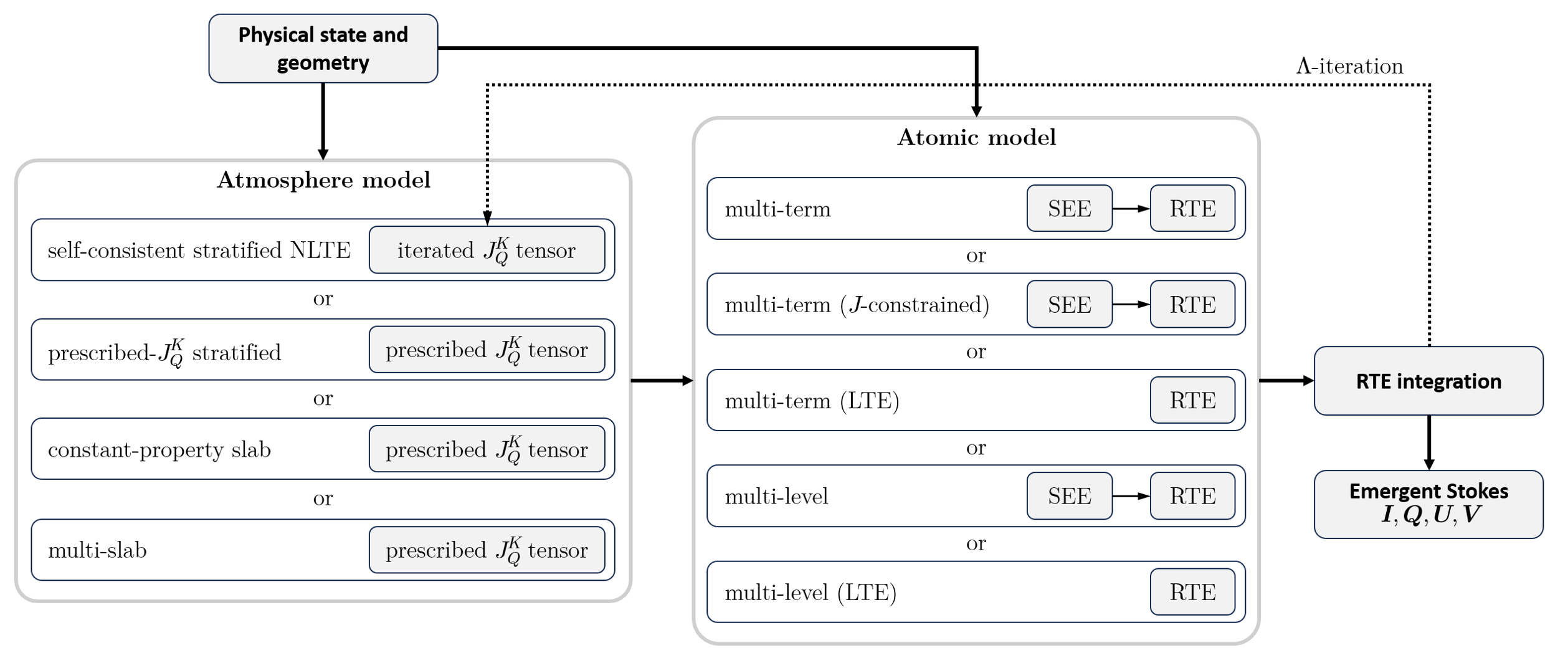}
  \caption{Data flow of a \solrat\ synthesis. The dashed arrow marks the
    self-consistent $\Lambda$-iteration update of $J^K_Q$.}
  \label{fig:architecture}
\end{figure}

Figure~\ref{fig:architecture} shows the \solrat\ data flow.
A synthesis starts from an atomic model that specifies the levels, the radiative transitions,
and the atomic description.
The atmosphere supplies the local thermodynamic state, the magnetic field vector,
and the velocity at each point along the ray.
The solution of the SEE returns the statistical tensors,
and the RTE stage turns them into the propagation matrix and the emission vector at every frequency.
The formal solution then advances the Stokes vector through the medium and delivers the emergent profile.
In the self-consistent case the emergent radiation updates the radiation field driving the SEE,
and the SEE is solved again.

The radiation field enters the SEE through the irreducible radiation tensor $J^K_Q$.
Its components are angular moments of the Stokes vector weighted by the polarization tensor $\mathcal T^K_Q$,
\begin{equation}
  J^K_Q(\nu) = \sum_{i=0}^{3}\oint\frac{\mathrm{d}\Omega}{4\pi}\,
    \mathcal T^K_Q(i,\hat{\bm\Omega})\,I_i(\nu,\hat{\bm\Omega})\,.
  \label{eq:JKQ}
\end{equation}
The tensor is evaluated in the reference frame used by the SEE after the geometric rotations described in
Appendix~\ref{sec:conventions}.

The SEE is solved for the statistical tensors of the selected atomic description.
In the multi-term case the unknowns include the inter-$J$ pairs $\rho^K_Q(J,J')$ inside each term; in the multi-level
case they reduce to one tensor $\rho^K_Q(\alpha,J)$ per level.
Both descriptions use the same operator classes: magnetic precession and fine-structure evolution, radiative pumping,
spontaneous and stimulated emission, and the corresponding relaxation terms.
In the incomplete Paschen-Back regime the multi-term density matrix is transformed from the $J$ basis to
the field-dependent Paschen-Back basis before the linear system is assembled and back after it is solved.
The explicit SEE structure is given in Appendix~\ref{sec:app_equations}.

By default the SEE describes a purely radiative scattering atom.
The computed scattering polarization is then an upper limit.
\solrat\ provides an optional parametrized collisional term (LL04 Sect.~7.13) attached to a model
in either atomic description.
Per radiative transition an inelastic de-excitation rate $C_{ul}$ is supplied.
The excitation rate $C_{lu}$ follows from the detailed-balance relation (LL04 eq.~7.98),
so that in the collision-dominated limit the populations relax to their Boltzmann values
and the line thermalizes toward LTE.
Per level, elastic depolarizing rates $D^{(K)}$ with $K\geq1$ (LL04 eq.~7.102) relax the alignment
and orientation while preserving the population, bridging the scattering and LTE limits.
The rate-transfer multipoles are represented by the population rate, $C^{(K)}=C^{(0)}$.
In LTE mode the statistical tensor is filled analytically from thermal populations and the SEE solve is bypassed.
The line shape and Stokes $V$ reflect the propagation matrix, including Paschen-Back corrections at very strong fields.

\begin{table*}
  \centering
  \caption{Capability matrix of the main atomic configurations. ``N/A''
    marks a capability that carries no meaning for that configuration.}
  \label{tab:capability}
  \begin{tabular}{lcccc}
    \hline\hline
    Feature & Multi-term & Multi-term LTE & Multi-level & Multi-level LTE \\
    \hline
    DELO transfer                                    & yes        & yes        & yes                & yes \\
    Prescribed-$J^K_Q$ atmospheres                   & yes        & yes\tablefootmark{a} & yes & yes\tablefootmark{a} \\
    Self-consistent stratified non-LTE atmosphere    & yes        & N/A        & yes                & N/A \\
    Empirical Land\'e factor                         & no\tablefootmark{b} & no\tablefootmark{b} & yes & yes \\
    Atomic level polarization                        & yes        & no         & yes                & no \\
    Parametrized collisions                          & yes        & N/A        & yes                & N/A \\
    Paschen-Back diagonalization                     & yes        & yes        & N/A                & N/A \\
    Inter-$J$ coherences                             & yes        & no         & no                 & no \\
    $J$-constrained RTE                              & yes        & yes        & N/A                & N/A \\
    \hline
  \end{tabular}
  \tablefoot{
    \tablefoottext{a}{In the LTE limit the statistical tensors are filled
    from the local temperature, so the prescribed $J^K_Q$ is ignored.}
    \tablefoottext{b}{The Land\'e factor follows from the LS term. The
    magnetic sensitivity can be adjusted toward a measured value with the
    $\xi$ scaling of the anomalous spin contribution.}
  }
\end{table*}

The Stokes parameters evolve along a ray path $s$ through an active medium according to the polarized RTE
\begin{equation}
  \frac{\mathrm{d} \bm{\mathrm{I}}}{\mathrm{d} s}
    = -\mathbf{K}\,\bm{\mathrm{I}} + \bm{\mathrm{\varepsilon}}\,,
  \label{eq:rte_vector}
\end{equation}
where $\mathbf{K}$ is the $4\times 4$ propagation matrix and $\bm{\varepsilon}$ is the emission vector.
The propagation matrix includes both dichroism ($\eta_I,\eta_Q,\eta_U,\eta_V$) and anomalous
dispersion ($\rho_Q,\rho_U,\rho_V$) and has a form
\begin{equation}
  \setlength\arraycolsep{3pt}
  \mathbf{K} =
  \begin{pmatrix}
    \eta_I  & \eta_Q   & \eta_U  & \eta_V  \\
    \eta_Q  & \eta_I   & \rho_V  & -\rho_U \\
    \eta_U  & -\rho_V  & \eta_I  & \rho_Q  \\
    \eta_V  & \rho_U   & -\rho_Q & \eta_I
  \end{pmatrix}
  \,.
  \label{eq:K_matrix}
\end{equation}
The diagonal $\eta_I$ is the absorption coefficient, the $\eta_{Q,U,V}$ couple the intensity to the polarized Stokes
parameters and carry the dichroism, and the $\rho_{Q,U,V}$ mix the polarized components among
themselves through the magneto-optical effects.
The emission vector $\bm\varepsilon$ sets the polarized source function $\mathbf S=\mathbf K^{-1}\bm\varepsilon$.
Given the SEE solution $\rho^K_Q$, the RTE constructs the propagation matrix $\mathbf K$
and the emission vector $\bm\varepsilon$.
Lower level statistical tensors enter the absorption and dichroism terms,
while upper level tensors enter the emissivity.
In the multi-term description, the Paschen-Back eigenvectors and field-shifted component frequencies enter
these coefficients, so the transfer coefficients vary nonlinearly with $B$ outside the linear Zeeman regime.
The explicit coefficient structure is given in Appendix~\ref{sec:app_equations}.

The complex line profile for each magnetic component is the Faraday-Voigt function
$\phi_q(\nu)=H(a,u_q)+\mathrm i\,L(a,u_q)$, where $H$ is the Voigt function, $L$ its dispersion counterpart,
$u_q=(\nu-\nu_0-q\nu_Z)/\Delta\nu_D$ the reduced frequency relative to the magnetically shifted line center,
$a$ the damping parameter, and $\Delta\nu_D$ the Doppler width.
Here $\nu$ is the observer-frame frequency, $\nu_0$ the zero-field line center frequency, $\nu_Z$ the
Zeeman shift, and $q$ the index of the magnetic component.
Input air wavelengths are converted to vacuum wavelengths with the Edl\'en formula \citep{edlen1966refractive}
before the line center frequency is formed.
\solrat\ evaluates this complex profile with the rational approximation of the Huml\'i\v{c}ek
algorithm \citep{humlivcek1982optimized}.
The real part gives the absorption and emission and the imaginary part the anomalous-dispersion
coefficients of Eq.~\eqref{eq:K_matrix}.
The Doppler width contains thermal and microturbulent broadening,
$\Delta\nu_D=(\nu_0/c)\sqrt{2k_BT/m_{\rm atom}+\xi_t^2}$, and the line-of-sight velocity shifts all
components by the same amount before the magnetic splitting is applied, so the dispersion terms respond
to the same thermal, turbulent, and damping parameters as the absorptive terms.

The transfer equation is evaluated through atmosphere classes that specify how the local plasma state and,
when needed, the prescribed radiation tensor vary along the ray.
The simplest atmosphere is a single plane-parallel slab of uniform physical properties
(temperature, magnetic field vector, turbulent and line-of-sight velocity, line optical depth $\tau_{\rm line}$,
and continuum-to-line ratio).
Constant-property slabs and multi-slab atmospheres can take a prescribed radiation tensor $J^K_Q$ as an
input to the SEE.
Within the slab the source function vector $\mathbf S=\mathbf K^{-1}\bm\varepsilon$ is constant and the
transfer equation~\eqref{eq:rte_vector} admits the closed-form DELO solution \citep[][see also \citealp{d2024impact}
for a comparison of DELO variants]{degl1985solution}
\begin{equation}
  \mathbf I_{\rm out} =
    \mathbf S
    + e^{-\mathbf K\tau}\bigl[\mathbf I_{\rm in} - \mathbf S\bigr]\,,
  \label{eq:DELO}
\end{equation}
where $e^{-\mathbf K\tau}$ is the matrix exponential.
The optical-depth parametrization normalizes the transfer coefficients by the line center opacity,
so the magnetic and polarimetric response can be explored separately from the absolute absorber density.

The multi-slab atmosphere chains several constant-property slabs, feeding the emergent Stokes vector of one
layer as the incident vector of the next.
Each spectral line has its own optical depth in each slab, so lines of different oscillator strength and
Land\'e factor probe the piecewise-constant structure differently.
This construction leaves hydrostatic balance and continuity across slab boundaries to the chosen input model.
A height-stratified atmosphere instead uses a depth grid on which the thermodynamic state, magnetic field,
velocity, density, broadening, and continuum opacity may vary.
In the prescribed-radiation version of this height-stratified atmosphere, the radiation tensor may also vary with
height as $J^K_Q(z)$ while still being supplied externally.

\solrat\ supports two ways of obtaining $J^K_Q$.
The prescribed-radiation atmosphere classes take it as an explicit input to the SEE.
This setup is appropriate for optically thin slabs, externally illuminated slabs, and controlled tests of
atomic-polarization physics, where the atomic problem is separated from the construction of the illumination.
The input can be a flat Planckian field or the parametrized Allen-continuum illumination \citep{cox2015allen} used,
for example, by \textsc{Hazel2} \citep{asensio2008advanced}.
For this Allen-type illumination the incident field is taken to be cylindrically symmetric about the local vertical.
Only the $Q=0$ moments survive,
\begin{equation}
  J^0_0(\nu) = \frac{1}{4\pi}\oint I\,\mathrm{d}\Omega\,,
  \label{eq:J00}
\end{equation}
and the quadrupole moment is
\begin{equation}
  J^2_0(\nu) = \frac{1}{4\pi}\oint
    \frac{3\cos^2\vartheta-1}{2\sqrt2}\,I\,\mathrm{d}\Omega\,,
  \label{eq:J20}
\end{equation}
where $\vartheta$ is the polar angle of the ray.
The prescribed field is then specified by the photon occupation number $n$ and by the anisotropy factor
$w=J^2_0/J^0_0$ at each transition frequency.
In the self-consistent height-stratified atmosphere model, $J^K_Q$ is instead reconstructed from the internal
radiation field, so that its spatial distribution is consistent with the SEE and RTE.

Prescribing $J^K_Q$ is the standard approximation behind many Hanle--Zeeman slab models.
When photons emitted inside the medium contribute appreciably to the local angular moments of the radiation field,
the coupled fixed-point problem must be solved:
\begin{equation}
  \rho^K_Q(z) = \mathcal E\!\left[J^K_Q(z)\right],
  \label{eq:see_map}
\end{equation}
\begin{equation}
  J^K_Q(z) = \Lambda^K_Q\!\left[\rho^K_Q(z')\right]\,,
  \label{eq:radiation_map}
\end{equation}
where $\mathcal E$ denotes the statistical equilibrium solve and $\Lambda^K_Q$ the angular and
frequency moments of the polarized transfer solution.
The prescribed-radiation atmospheres apply Eq.~\eqref{eq:see_map}, with $J^K_Q$ supplied as a fixed input to the SEE.
The self-consistent stratified atmosphere applies Eqs.~\eqref{eq:see_map} and \eqref{eq:radiation_map} on the same depth grid.

\solrat\ solves this one-dimensional radiation field problem in its height-stratified atmosphere class.
Physical parameters -- temperature, absorber number density, the magnetic field vector, microturbulence, Voigt damping,
and the macroscopic-velocity vector -- can vary continuously with geometric height on a user-defined depth grid,
supplied as a constant, an array, or a callable of height.
The opacity scale is carried by the local number density $N(z)$, so the optical depth follows
from $N(z)$ and the geometry.
A $\Lambda$-iteration alternates between Eqs.~\eqref{eq:see_map} and \eqref{eq:radiation_map}.
Given the current $\rho^K_Q(z)$, the polarized transfer equation is solved by a DELO evolution-operator step along
a set of rays -- either a first-order (piecewise-constant source) or a second-order (DELO-linear) scheme.
The emergent Stokes distribution is projected onto $\mathcal T^K_Q(i,\Omega)$, averaged in frequency
(with each transition's own normalized absorption profile) and in angle to reconstruct $J^K_Q(z)$,
and the SEE is re-solved per depth.
The iteration stops when $\max|\Delta\rho|$ falls below a tolerance.
An optional error estimate can instead stop on the inferred distance to the fixed point, computed from the recent
geometric decay of this residual.

The angular integral is evaluated with a double-Gauss quadrature in $\mu=\cos\vartheta$, using a separate Gauss-Legendre rule on each hemisphere and a uniform azimuthal sampling.
Splitting at $\mu=0$ places the surface radiation field kink on a subinterval boundary.
The split restores spectral convergence and removes a low bias in the surface anisotropy,
and hence in $\rho^2_0/\rho^0_0$.
A vertical velocity gradient enters through the Doppler-shifted absorption profile
evaluated at the local velocity projection of each ray.
A gray continuum with an LTE source is added per depth \citep{khan2006stellar}.
A tangential line of sight ($\mu\to0$) is handled in the Eddington-Barbier limit $I(0)=S(\tau{=}0)$,
and returns the surface source function for the tangential ray.
Optional Ng acceleration \citep{ng1974hypernetted} extrapolates the last four (or more) density matrix
iterates to their fixed point, preserving the trace normalization and cutting the iteration count.

The capabilities of the atomic configurations are summarized in Table~\ref{tab:capability}.
The rows separate the physical content that each configuration carries.
Fine-structure coherences and exact diagonalization of the magnetic Hamiltonian are
available in the multi-term configurations. Atomic level polarization and the
self-consistent radiation field are available whenever the equilibrium is solved, and
the LTE variants replace that solve with thermal populations. The prescribed-radiation
atmospheres and the transfer solver are shared by every configuration, so a change of
atomic description leaves the propagation untouched.

\section{Implementation}

\solrat\ is written in Python~3.
Its numerical and symbolic core is based on NumPy \citep{harris2020array}, SciPy \citep{virtanen2020scipy},
and SymPy \citep{meurer2017sympy}, and the figures are produced with Matplotlib \citep{hunter2007matplotlib}.
Pandas \citep{mckinney2010data} is used for one-time assembling of the dataframes and debugging views.
These packages provide the array representation, dense linear algebra, interpolation, special functions, and symbolic
angular momentum factors used by the code.
Computationally heavy operations are expressed as vectorized array operations and dense linear solves, so the numerical
work is passed to the compiled libraries behind NumPy and SciPy.
With this dependency stack, the code can be run on Windows, Linux, and macOS.

The implementation separates the atomic description from the atmosphere and formal-transfer drivers.
The same constant-property slab, multi-slab, and height-stratified atmosphere classes can therefore be used with either
the multi-term or the multi-level description.
Changing the atomic description changes the SEE and transfer-coefficient construction, while the atmospheric setup and
formal solution remain unchanged.
The built-in atomic models used in the examples are specified through the same term, level, and transition
registration used for additional atoms, so they also serve as templates for extending the atomic data.
In both descriptions the SEE is a linear time-independent system $\mathbf A\,\bm x=\mathbf 0$ closed by the
trace-normalization constraint $\sum_{\rm levels}\sqrt{2J+1}\,\rho^0_0=1$ for the total population.
\solrat\ eliminates the first row and column, performs a dense linear solve, and re-applies the normalization.
The size of the system follows from the selected atomic description and from the multipoles retained on each level or term.

The internal summation engine keeps each SEE rate and each RTE coefficient close to the corresponding equation.
Every constrained sum is represented as a columnar table of indices and factors, including Wigner symbols, Einstein
coefficients, geometric tensors, radiation tensors, and statistical tensors.
This representation localizes changes to the formal description:
for example, a different collisional parametrization
can be introduced by adding or replacing rate factors while the atmosphere and transfer drivers are retained.
The same mechanism can restrict the multi-term RTE to a selected $J\to J'$ branch while keeping the full term in
the Hamiltonian and SEE, which isolates one observed branch from neighboring fine-structure branches.
The summation engine also supports staged caching.
For a fixed atomic model, atom-specific factors such as angular momentum coefficients are
precomputed and reused in repeated syntheses of the same spectral line.
For a fixed atmosphere and geometry, the factors that depend on the local physical conditions are cached while
the statistical tensors change during a self-consistent iteration.
Appendix~\ref{sec:app_engine} describes the engine and the staged caching in more detail.

\section{Profile sensitivity to the atomic description}

\begin{figure}
  \centering
  \includegraphics[width=0.9\linewidth]{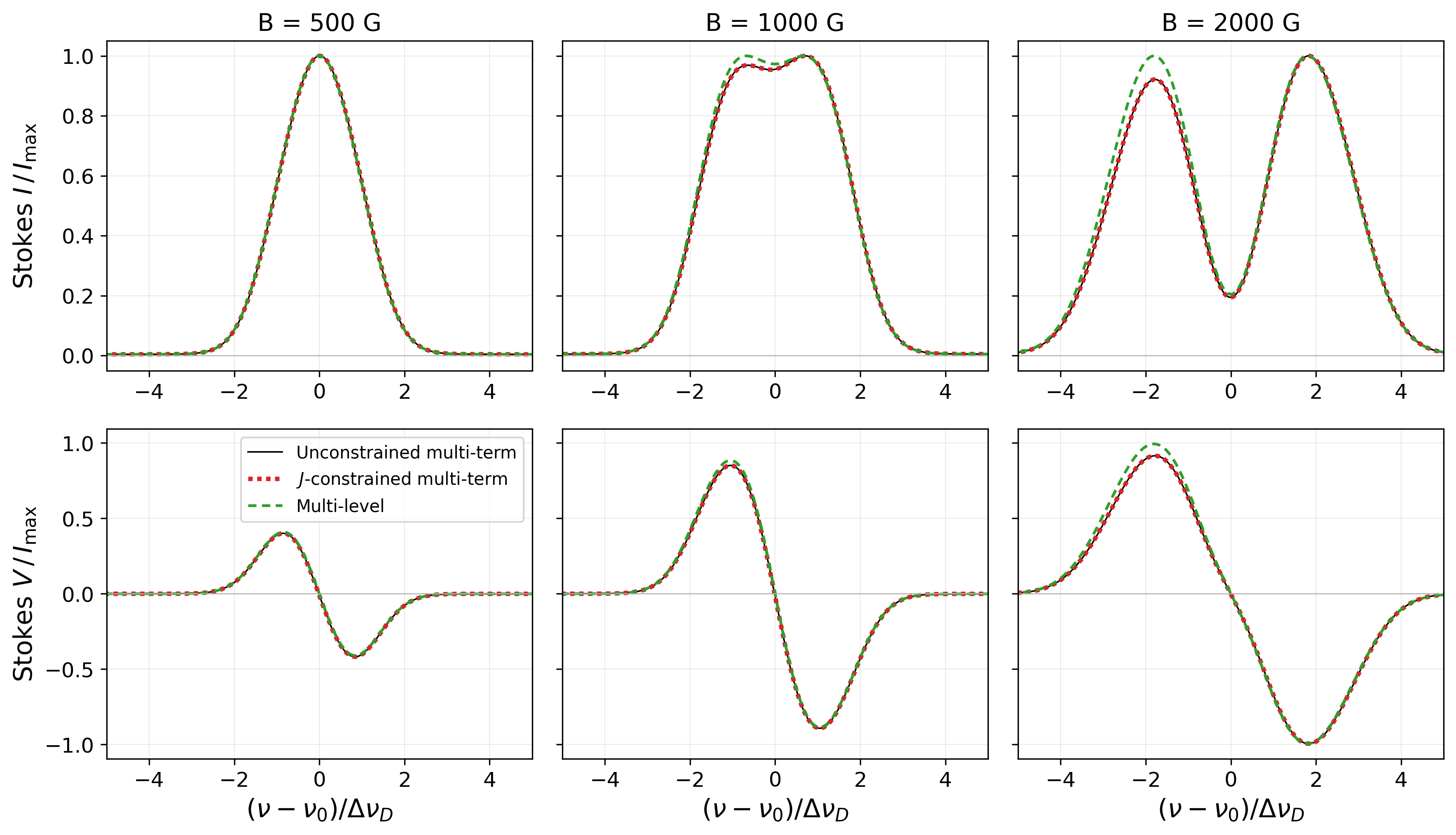}
  \caption{Second-order Zeeman comparison on the
    $^4P_{5/2}\to{}^4S_{3/2}$ branch. Columns show $B=500$, $1000$, and
    $2000$~G. Rows show Stokes $I/I_{\max}$ and Stokes $V/I_{\max}$. Solid lines:
    unconstrained multi-term. Dotted lines: $J$-constrained multi-term.
    Dashed lines: multi-level.}
  \label{fig:mt_ml_zeeman}
\end{figure}

The multi-term and multi-level descriptions can produce different Stokes profiles even when the fine-structure
splitting significantly exceeds the Zeeman splitting.
To illustrate this regime, Fig.~\ref{fig:mt_ml_zeeman} uses a synthetic $^4P\to{}^4S$ multiplet.
The selected transition is the $^4P_{5/2}\to{}^4S_{3/2}$ branch, and the neighboring upper term fine-structure
branches are placed several Doppler widths away (see Appendix~\ref{sec:app_applications} for details).
Here, unconstrained multi-term means that all allowed fine-structure branches contribute to the synthesis, whereas
$J$-constrained multi-term keeps the same multi-term atomic structure but restricts the RTE to the selected
$J\to J'$ branch.
In the multi-level description the magnetic sublevels follow linear Zeeman shifts, and the Zeeman component
strengths remain fixed.
In this field range the multi-term component positions remain close to the multi-level splitting, but the magnetic
field admixes neighboring $J$ values at fixed $M$ and redistributes the component strengths.
The Stokes profiles can therefore begin to differ through the intensity of the magnetic components before their
frequency positions show a large separation.
The onset field scales with the smallest fine-structure separation participating in the observed component,
$B\sim\Delta E_{\rm FS}/(g\mu_B)$, and the off-diagonal $\langle J\pm1,M|\hat S_z|J,M\rangle$ elements make
high-spin terms sensitive test cases.
In the core window shown in Fig.~\ref{fig:mt_ml_zeeman}, the unconstrained multi-term and $J$-constrained
multi-term curves nearly coincide because the satellite branches lie outside the plotted interval.
The multi-level curve starts to deviate mainly through the amplitudes of Stokes $I$ and $V/I_{\max}$, consistent
with fixed Zeeman component strengths in the multi-level description and field-dependent strength redistribution
in the multi-term description.

\begin{figure}
  \centering
  \includegraphics[width=0.5\linewidth]{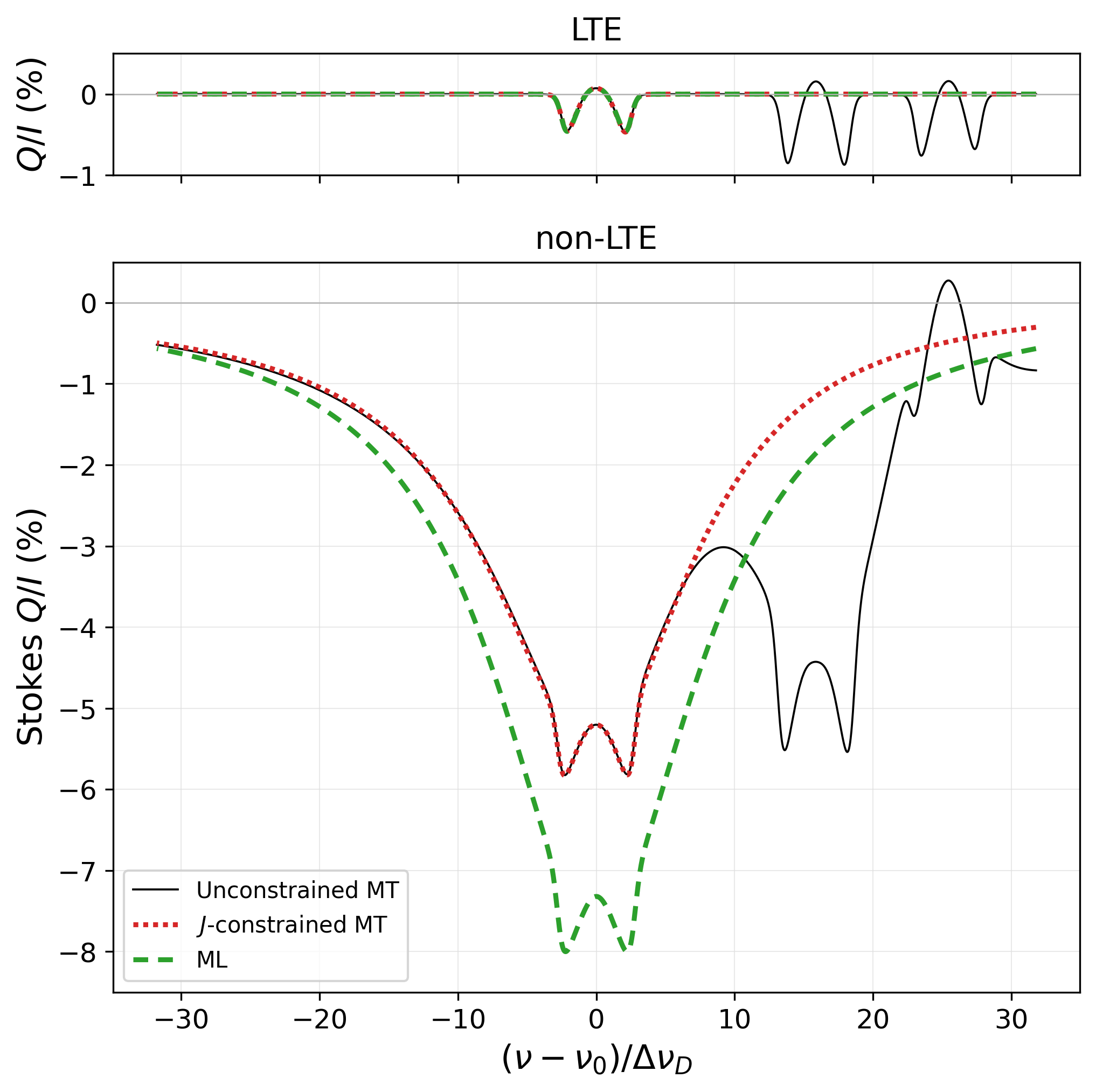}
  \caption{LTE (top) and non-LTE (bottom) Stokes $Q/I$ in percent for the
    $^4P_{5/2}\to{}^4S_{3/2}$ line at $B=50$~G. Solid lines:
    unconstrained multi-term. Dotted lines: $J$-constrained multi-term.
    Dashed lines: multi-level.}
  \label{fig:mt_ml_scattering}
\end{figure}

Figure~\ref{fig:mt_ml_scattering} uses the same synthetic multiplet at weak field ($B=50$~G) and prescribed
Allen-continuum illumination at $h=30''$.
In LTE, all three curves agree across the central region, roughly within ten Doppler widths of line center.
Farther from line center, the unconstrained multi-term curve separates because it is the only synthesis that
includes the neighboring fine-structure satellite branches.
In non-LTE, the unconstrained and $J$-constrained multi-term curves agree across the central region, roughly within
seven Doppler widths, and the unconstrained multi-term curve again separates in the broader wings.
The multi-level non-LTE profile has nearly the same shape as the $J$-constrained multi-term profile, but its
amplitude is larger in magnitude: the minimum plotted value is about $Q/I=-8\%$ for the multi-level description
and about $Q/I=-6\%$ for the multi-term descriptions.
At this weak field this difference reflects the inter-$J$ coherences retained by the multi-term description and
omitted in the multi-level one.

These two examples illustrate two important ways in which the atomic description enters the profile, beyond the
large-field departure of the Zeeman splitting from the linear regime.
The strong-field comparison isolates the redistribution of magnetic component strengths in the observed-line core.
The weak-field comparison shows that, although the LTE multi-term and multi-level profiles are nearly identical
near line center, non-LTE alignment can make their difference measurable even when the field is weak.
The close agreement between the unconstrained and $J$-constrained multi-term curves in the line core also shows
when the constrained calculation is a useful approximation: the selected branch can be isolated if the neighboring
fine-structure components do not overlap the spectral region of interest.

\section{Validation and benchmarks}

The validation set combines local numerical checks, code-to-code comparisons, and published radiative transfer
benchmarks.
Table~\ref{tab:validation_battery} lists the corresponding reproducibility checks.
The analytic rows test isolated pieces of the synthesis, including the Voigt profile, formal integration, angular
factors, and limiting magnetic-polarization formulae.
The benchmark rows compare complete syntheses with \textsc{Hazel2}, with the self-consistent resonance-scattering
benchmark of \citet{trujillo1999iterative}, hereafter TM99, and with the thermalization benchmark of
\citet{avrett1965noncoherent}, hereafter AH65.
The last column reports the absolute root-mean-square (RMS) difference printed by the corresponding reproduction
script.
The Voigt-profile row compares the Huml\'i\v{c}ek approximation used in \solrat\ with a SciPy
Faddeeva-function evaluation \citep{virtanen2020scipy}.
These checks validate the angular algebra, line profile evaluation, formal transfer, and self-consistent
radiation field iteration under controlled conditions.
Appendix~\ref{sec:app_applications} gives the benchmark configurations and additional setup details.

\begin{table*}
  \centering
  \caption{Reproducibility checks and their absolute agreement metrics.}
  \label{tab:validation_battery}
  \begin{tabular}{p{0.25\linewidth}p{0.26\linewidth}p{0.30\linewidth}p{0.12\linewidth}}
    \hline\hline
    Check & Compared against & Quantity compared & Absolute RMS \\
    \hline
    Huml\'i\v{c}ek Voigt approximation & SciPy Faddeeva function, LL04 Sect.~5.4
      & Voigt profile $H(a,v)$ & $5\cdot10^{-6}$ \\
    DELO formal solution & Dense finite-difference integration
      & Emergent Stokes vector & $5\cdot10^{-4}$ \\
    Stratified prescribed-$J^K_Q$ transfer & Analytic Unno-Rachkovsky solution, LL04 Sect.~9.8
      & Stokes $I$, $Q/I$, $U/I$, $V/I$ & $8\cdot10^{-7}$ \\
    Multi-term vs multi-level ($S=0$) & Independent multi-level atom for the same singlet--singlet transition
      & Stokes $I$, $Q/I$, $U/I$, $V/I$ & $1\cdot10^{-14}$ \\
    Single-scattering polarization & Rayleigh angular factor, LL04 Sect.~10.2
      & Normalized Stokes $Q/I$ & $9\cdot10^{-5}$ \\
    Hanle depolarization factor & Analytic Hanle factor, LL04 Sect.~10.3
      & $|\rho^2_2|/|\rho^2_2|_{B=0}$ & $9\cdot10^{-4}$ \\
    \textsc{Hazel2} comparison, on-disc profile & \textsc{Hazel2} He\,\textsc{i}~D$_3$ synthesis
      & Stokes $V/I$ at line-of-sight angle $60^\circ$, $|B|=8000$~G & $2\cdot10^{-4}$ \\
    \textsc{Hazel2} comparison, limb profile & \textsc{Hazel2} He\,\textsc{i}~D$_3$ synthesis
      & Stokes $Q/I$ at line-of-sight angle $90^\circ$, $|B|=1000$~G & $5\cdot10^{-5}$ \\
    Self-consistent scattering alignment & Digitized TM99 Fig.~8\tablefootmark{a}
      & $\rho^2_0/\rho^0_0(\tau)$ for $\delta^{(2)}=1$
      & $8\cdot10^{-4}$ \\
    Self-consistent scattering alignment & Digitized TM99 Fig.~8\tablefootmark{a}
      & $\rho^2_0/\rho^0_0(\tau)$ for $\delta^{(2)}=0.1$
      & $1.6\cdot10^{-3}$ \\
    Emergent scattering polarization & Digitized TM99 Fig.~10\tablefootmark{a}
      & $Q/I(\nu)$ for $\mu=0.1$ & $2.2\cdot10^{-4}$ \\
    Thermalization curve & Digitized AH65 Fig.~2\tablefootmark{a}
      & $S(\tau)/B$ for $\epsilon=10^{-2}$ & $3\cdot10^{-3}$ \\
    \hline
  \end{tabular}
  \tablefoot{
    \tablefoottext{a}{The quoted precision is limited by digitization of
    the published figure.}
  }
\end{table*}

\begin{figure}
  \centering
  \includegraphics[width=0.5\linewidth]{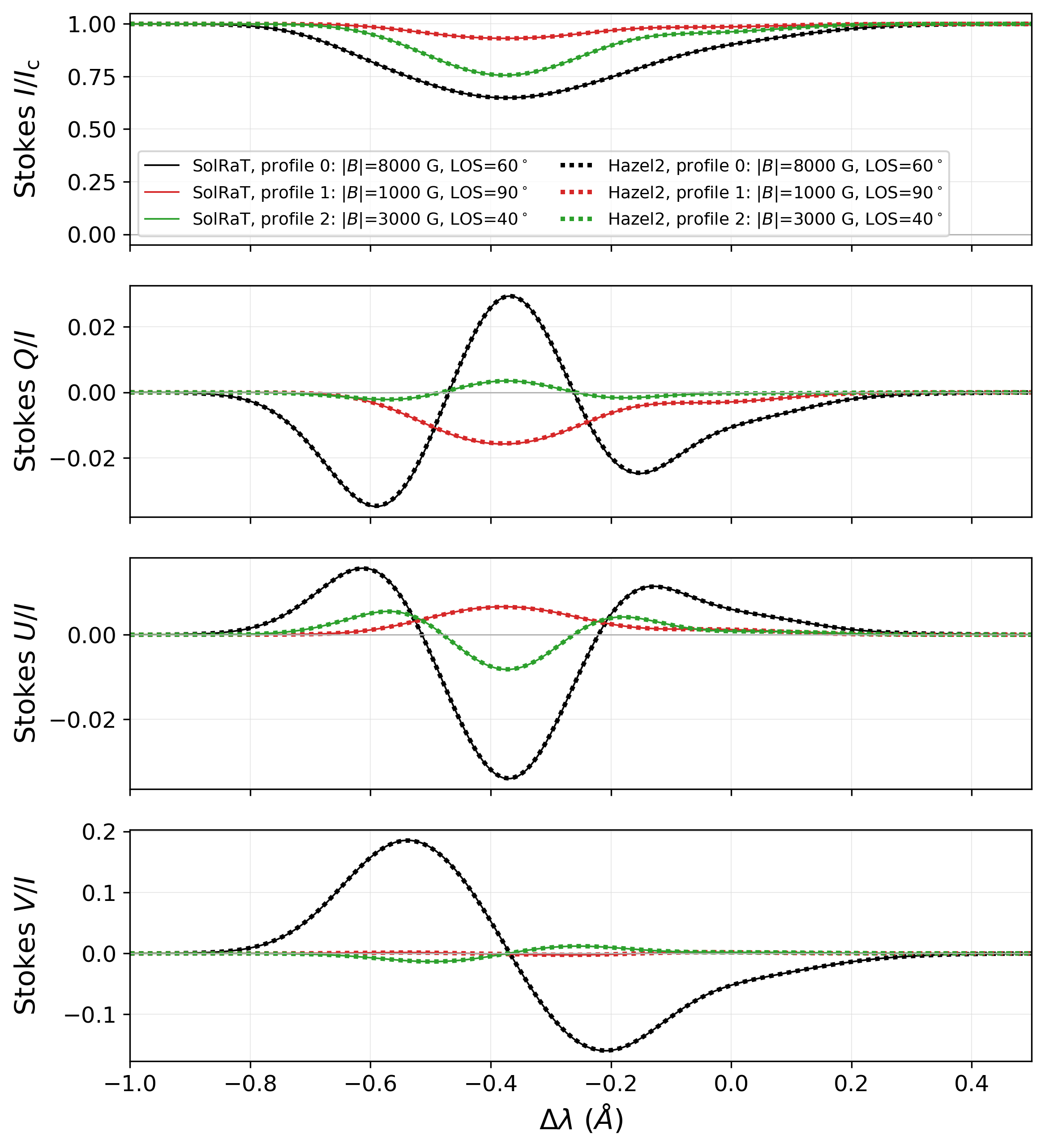}
  \caption{He\,\textsc{i}~D$_3$ Stokes profiles from \solrat\ (solid)
    and \textsc{Hazel2} (dotted). Color marks the slab configuration.
    Here $I_{\rm c}$ is the continuum intensity.}
  \label{fig:HeI_hazel}
\end{figure}

The He\,\textsc{i}~D$_3$ comparison with \textsc{Hazel2} uses constant-property slabs in their common domain of
validity (Fig.~\ref{fig:HeI_hazel}).
The on-disc configuration tests the circular polarization at strong field, where incomplete Paschen-Back mixing
shapes the profile.
The limb configuration tests the linear polarization at weak field, where scattering polarization and the Hanle
effect set the signal.
The incident continuum, prescribed anisotropic radiation tensor, and slab parameters are matched in each case.
The residual is not expected to reach machine precision because it accumulates differences from a set
of independent implementation choices.
In the zero-damping setup used here, \textsc{Hazel2} evaluates the Faraday-Voigt profile with its small-damping
Dawson-function branch, whereas \solrat\ uses the Huml\'i\v{c}ek complex-profile approximation throughout.
The two codes also place the LL04 frame rotation at different stages of the calculation: \textsc{Hazel2} solves the
SEE in the local vertical frame and rotates the density tensor before evaluating the RTE coefficients, whereas
\solrat\ rotates the radiation tensor into the magnetic field frame before solving the SEE.
The two codes nevertheless show a close agreement, as is evident from the overplotted profiles in Fig.~\ref{fig:HeI_hazel}.

\begin{figure}
  \centering
  \includegraphics[width=0.5\linewidth]{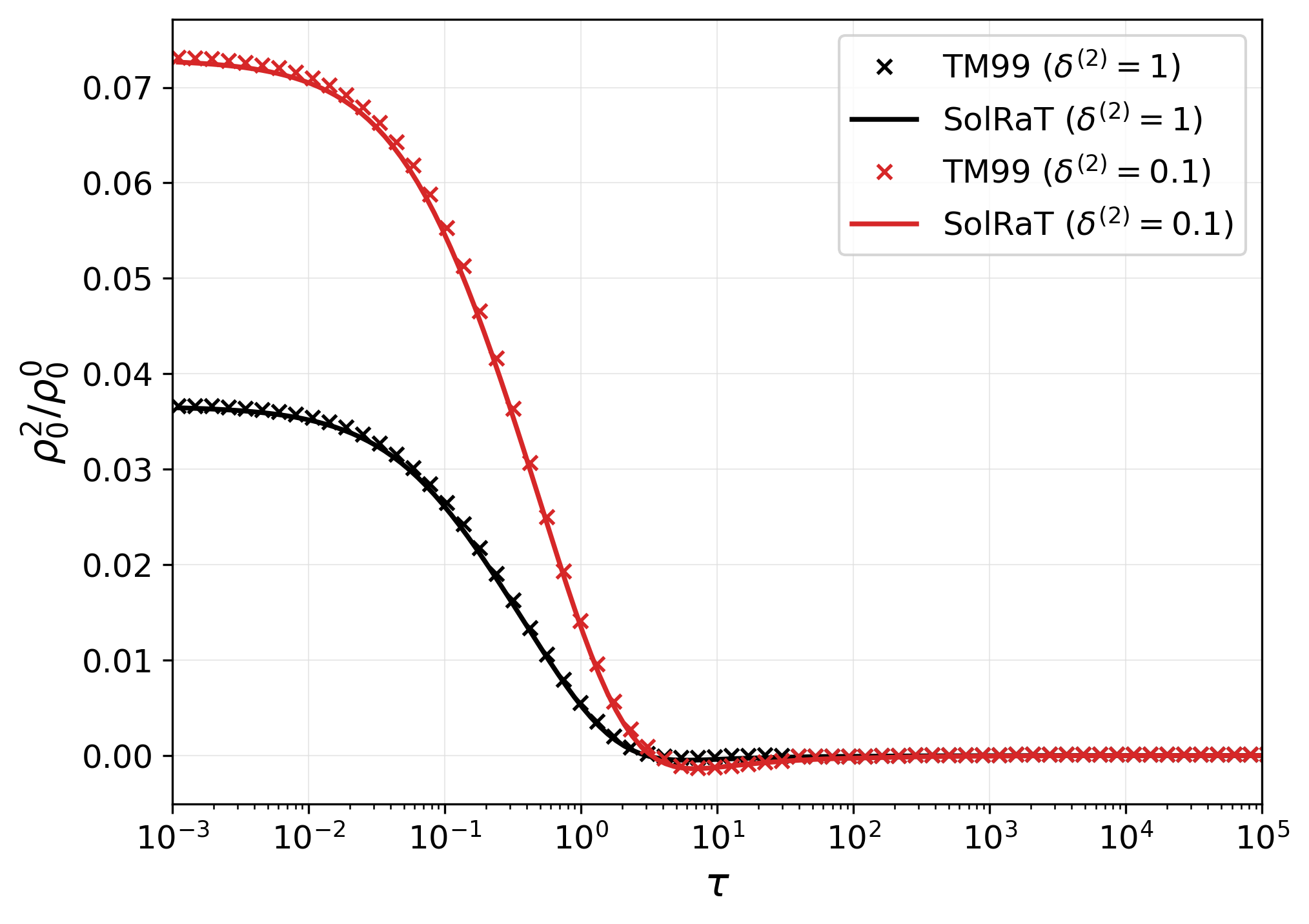}
  \caption{Upper level alignment in the TM99 slab benchmark. Curves show
    \solrat. Crosses show digitized TM99 Fig.~8 data for
    $\delta^{(2)}=1$ and $0.1$.}
  \label{fig:tm1999}
\end{figure}

The self-consistent solution of SEE and RTE is validated with the TM99 resonance line benchmark.
Figure~\ref{fig:tm1999} shows the converged upper level alignment $\rho^2_0/\rho^0_0(\tau)$ for a $J=0\to1$
two-level atom in an isothermal slab, reproducing the TM99 benchmark for two elastic depolarizing rates.
The dependence on the elastic depolarizing rate follows the TM99 result throughout the slab.
This comparison checks the self-consistent radiation field iteration used by the stratified non-LTE atmosphere.
Appendix~\ref{sec:app_applications} provides additional comparisons: the emergent TM99 $Q/I$ profile, the AH65
source-function thermalization benchmark, the stratified transfer solution against the Unno-Rachkovsky solution,
the magnetic coherence damping against the analytic Hanle factor, and the multi-term description against its
multi-level counterpart in the $S=0$ limit.

\section{Discussion}

The multi-term and multi-level syntheses agree in the formal limits where the two descriptions reduce to the same
problem, but they diverge when inter-$J$ coherences or Paschen-Back mixing affect the line.
In the comparisons shown here (Figs.~\ref{fig:mt_ml_zeeman} and \ref{fig:mt_ml_scattering}), the atmosphere,
radiation tensor, and transfer solver are kept fixed, so the profile differences can be traced to the atomic
description itself.
This makes the examples a controlled way to identify where the approximations enter the observables.

The two examples probe different parts of the atomic description.
In the strong-field example (Fig.~\ref{fig:mt_ml_zeeman}) the difference is set by the field, through the positions
and the strengths of the magnetic components, while in the weak-field example (Fig.~\ref{fig:mt_ml_scattering}) it
is set by the inter-$J$ coherences created under anisotropic illumination.
The practical choice between the descriptions follows from this, together with their cost: the multi-term system is
larger because it carries the inter-$J$ tensor components, and its rates contain the corresponding higher-order
angular algebra, whereas the multi-level description is cheaper and discards both the coherences and the
field-dependent redistribution of component strengths.
Running both descriptions in one code makes this trade-off explicit for a chosen line and atmosphere.

The strong-field corrections are becoming increasingly relevant due to recent reports of kilogauss and
super-kilogauss magnetic fields. Examples include sunspot light bridges \citep{castellanos2025superstrong},
line profile diagnostics of major flares \citep{lozitsky2018profiles,yakovkin2024altitude, lozitska2024unique}
and sunspots \citep{lozitsky2026observations}, anomalous flare broadening
\citep{gomez2026enigmatic}, and force-free flux-rope models of flare energy release supporting
superstrong magnetic fields \citep{solov2022force}.
The size of the correction is line-dependent, set mainly by the fine-structure separation, term spin, and effective
Land\'e factor. Lines with weak linear Zeeman sensitivity can be especially sensitive to the nonlinear Zeeman corrections.

The limitations of the present \solrat\ release warrant a more detailed discussion.
The radiation tensor is evaluated in the flat-spectrum and complete frequency redistribution (CRD) approximation.
Partial redistribution, important in the wings of strong resonance lines
\citep{alsina2018magneto, janett2021modeling}, would require a frequency-resolved radiation tensor.
In this regard, a promising framework for future implementation in \solrat\ was developed in
\citep{sowmya2014polarized, sampoorna2017polarized}, where the partial frequency redistribution is combined
with the multi-term atom supporting arbitrary magnetic field strengths.
Collisional rates in \solrat\ are parametrized, with an inelastic de-excitation rate per transition and elastic
depolarizing rates per level supplied as input, and with the rate transfer multipoles represented by the population
rate.
Detailed $K>0$ rates derived from atom-collider cross sections \citep{derouich2003semi} and complete
irreducible-tensor multi-term collisions \citep{belluzzi2013isotropic} remain outside this parametrization.
Finally, the self-consistent radiation field iteration is one-dimensional and plane-parallel, so geometries in which
horizontal structure shapes the anisotropy of the radiation field stay outside the scope of the present release.

Each of these restrictions can be lifted in future development, and the structure of the code accommodates the
corresponding work.
A frequency-resolved radiation tensor for PRD, accelerated operator splitting for optically thick lines
\citep{olson1986rapidly, rybicki1991accelerated, rybicki1992accelerated}, and full-azimuth quadratures for
non-plane-parallel geometries all act on the operator that maps the local statistical tensors to the radiation
tensor, which the height-stratified atmosphere already replaces while reusing the same atom-specific SEE and RTE
drivers.
Collisional rates derived from cross sections enter instead as additional rate factors inside the existing SEE
structure, and an inversion layer would be built on the same forward synthesis interface.
In each case the atomic descriptions and the transfer solver keep their present form, so the atmosphere and transfer
setup can still be held fixed while the atomic description is changed, and the physical origin of a profile
difference remains identifiable.

\section{Conclusions}

\solrat\ is introduced as an open-source forward-synthesis code for polarized spectral lines in the density-matrix
framework of LL04.
It treats atomic polarization, anisotropic radiation, Zeeman splitting, and incomplete or complete Paschen-Back
mixing within one interface, so that multi-term and multi-level descriptions can be applied to the same atmosphere
and radiation field.
The implementation is validated against analytic transfer and Hanle limits, the \textsc{Hazel2}
He\,\textsc{i}~D$_3$ synthesis, and published self-consistent scattering and thermalization benchmarks.

The multi-term and multi-level descriptions agree in the limits where the atomic descriptions become formally
equivalent, but they separate once the retained physics differs.
In the synthetic $^4P\to{}^4S$ example, the multi-term synthesis departs from the linear-Zeeman multi-level result
between $500$ and $2000$~G mainly through the strengths of the magnetic components.
At $50$~G under anisotropic illumination, the separation instead reflects the inter-$J$ coherences retained by the
multi-term description, which change the amplitude of the scattering polarization.

These comparisons show that a diagnostic based on a single atomic description can carry regime-dependent
systematic errors.
Running interchangeable atomic descriptions through the same synthesis setup makes those errors identifiable for a
chosen line, and provides a path toward scattering-polarization diagnostics applicable across multiple atomic
descriptions.

\section*{Data availability}

\solrat\ is free and open-source software. The source code, the built-in
atomic data, links to the online documentation, and the demonstration scripts that reproduce every figure in
this paper are available in the public repository
\url{https://github.com/yakovkinii/SolRaT}, under the git tag
\texttt{manuscript}.

\section*{Acknowledgements}
The authors declare no conflicts of interest.
\bibliographystyle{apalike}
\bibliography{main} 

\appendix
\renewcommand{\thesection}{Appendix~\Alph{section}}

\section{Conventions}
\label{sec:conventions}

The most important conventions for comparing \solrat\ results with other codes are summarized below.
Overall, \solrat\ adopts the LL04 conventions throughout.
Geometric quantities are expressed in the fixed reference frame of LL04, with the $z$ axis along
the outward atmosphere normal.
A propagation direction is $\hat{\bm\Omega}=(\sin\theta\cos\chi,\sin\theta\sin\chi,\cos\theta)$ with $\mu=\cos\theta$.
The line of sight is specified by its polar and azimuthal angles $(\theta,\chi)$, and the magnetic field and
macroscopic velocity by their own polar and azimuthal angles $(\theta_B,\chi_B)$ and $(\theta_v,\chi_v)$ in
the same frame.
The positive-$Q$ reference direction is fixed by the observer angle $\gamma$, $U$ completes a right-handed set,
and the Stokes parameters $I,Q,U,V$ together with the propagation matrix follow the
LL04 definitions.

With this convention, for a magnetic line the Stokes-$V$ absorption coefficient is
\begin{equation}
  \eta_V = \tfrac{1}{2}\,(\varphi_r - \varphi_b)\,\cos\theta_B\,,
  \label{eq:eta_V_sign}
\end{equation}
where $\varphi_b$ and $\varphi_r$ are the absorption profiles of the blue and red $\sigma$ components,
centered at $\nu_0+g\nu_L$ and $\nu_0-g\nu_L$ respectively, $\nu_L$ is the Larmor frequency,
$g$ the effective Land\'e factor, and $\theta_B$ is the angle between the magnetic field and the line of sight.
The magneto-optical coefficient $\rho_V$ follows the same pattern with the Faraday-Voigt dispersion profiles $\psi$.

The Doppler shift of the absorption profile uses the projection $v_{\rm los}=-\hat{\bm\Omega}\cdot\bm v$, with
$\hat{\bm\Omega}$ the photon propagation (toward-observer) direction.
With this sign, a flow with a component toward the observer gives $v_{\rm los}<0$ and a blueshift.
The Zeeman shift of a sublevel is $g\,\mu_B B\,M$.
The multi-term magnetic Hamiltonian matrix elements and the $\xi$ scaling are given explicitly in
Appendix~\ref{sec:app_hamiltonian}, with $\xi=1$ recovering strict LS coupling.

Three reference frames are involved in any line synthesis.
The prescribed solar radiation field is naturally expressed in the local-vertical frame.
The SEE is solved in the magnetic frame, where the magnetic-precession operator is diagonal in $Q$.
The Stokes parameters are defined in the line-of-sight frame.
\solrat\ therefore rotates the radiation tensor from the local-vertical frame to the magnetic frame before solving
the SEE, and then evaluates the RTE coefficients for the observer line of sight.

\section{Magnetic Hamiltonian}
\label{sec:app_hamiltonian}

Within a single LS term, and with $\bm B$ along the quantization axis, the sum of the spin-orbit and magnetic
Hamiltonians is tridiagonal in $J$ at fixed $M$.
In the $|\beta LSJM\rangle$ basis, \solrat\ uses
\begin{equation}
  H_{J,J}
  = E_J + \mu_B B M
    \left[
      1 + \frac{\xi}{2}
      \frac{J(J+1)+S(S+1)-L(L+1)}{J(J+1)}
    \right]
  \label{eq:app_hamiltonian_diag}
\end{equation}
for $J>0$.
For the only $J=0$ sublevel, $M=0$ and $H_{0,0}=E_0$.
The neighboring off-diagonal element is
\begin{equation}
  \begin{aligned}
  H_{J-1,J}=H_{J,J-1}
  &= -\frac{\xi\,\mu_B B}{2J} \\
  &\quad \times \biggl[
    (J+S+L+1)(J-S+L)(J+S-L) \\
  &\quad \times (-J+S+L+1)\,
    \frac{J^2-M^2}{(2J+1)(2J-1)}
  \biggr]^{1/2}.
  \end{aligned}
  \label{eq:app_hamiltonian_offdiag}
\end{equation}
where scalar $\xi$ is an optional scale applied to the anomalous spin contribution.
The matrix is diagonalized numerically at each field strength, giving the Paschen-Back eigenvalues and
the eigenvectors $C^j_J(\beta LS,M)$, spanning the linear Zeeman, incomplete, and complete Paschen-Back regimes
(Fig.~\ref{fig:paschen_back}).
By default, \solrat\ uses $\xi=1$, for which Equations~\eqref{eq:app_hamiltonian_diag} and
\eqref{eq:app_hamiltonian_offdiag} recover the
standard LS-coupling matrix elements.
Setting $\xi\neq1$ tunes the magnetic sensitivity of the term, which can be relevant, for example,
for the Fe\,\textsc{i}~5434.5~\AA\ line synthesis \citep{landi1982effective}.

\section{Explicit expressions}
\label{sec:app_equations}

For a multi-term atom the SEE is written for the irreducible statistical tensor of each term.
The component $\rho^K_Q(J,J')$ carries the multipole rank $K$, projection $Q$, and the pair of fine-structure
levels $J,J'$ inside the term.
The equation has the form
\begin{equation}
  \begin{aligned}
  \frac{\mathrm d}{\mathrm d t}\rho^K_Q(J,J')
  ={}& -2\pi\mathrm i \sum_{K'Q',\,J''J'''}
       N(JJ'KQ,\,J''J'''K'Q')\,\rho^{K'}_{Q'}(J'',J''') \\
  &+ \sum_{\beta_\ell L_\ell,\,J_\ell J'_\ell K_\ell Q_\ell}
       T_A(JJ'KQ,\,J_\ell J'_\ell K_\ell Q_\ell)\,
       \rho^{K_\ell}_{Q_\ell}(J_\ell,J'_\ell) \\
  &+ \sum_{\beta_u L_u,\,J_u J'_u K_u Q_u}
       \bigl[T_E+T_S\bigr](JJ'KQ,\,J_u J'_u K_u Q_u)\,
       \rho^{K_u}_{Q_u}(J_u,J'_u) \\
  &- \sum_{K'Q',\,J''J'''}
       \bigl[R_A+R_E+R_S\bigr](JJ'KQ,\,J''J'''K'Q')\,
       \rho^{K'}_{Q'}(J'',J''')\,.
  \end{aligned}
  \label{eq:SEE_MT}
\end{equation}

For a multi-level atom each level $\alpha J$ carries a single statistical tensor, and the SEE becomes
\begin{equation}
  \begin{aligned}
  \frac{\mathrm d}{\mathrm d t}\rho^K_Q(\alpha J)
  ={}& -2\pi i\,\nu_L\,g_{\alpha J}\,Q\,\rho^K_Q(\alpha J) \\
  &+ \sum_{\alpha_l J_l, K_l Q_l}
       T_A(\alpha J K Q,\alpha_l J_l K_l Q_l)\,\rho^{K_l}_{Q_l}(\alpha_l J_l) \\
  &+ \sum_{\alpha_u J_u, K_u Q_u}
       \bigl[T_E + T_S\bigr](\alpha J K Q,\alpha_u J_u K_u Q_u)\,
       \rho^{K_u}_{Q_u}(\alpha_u J_u) \\
  &- \sum_{K' Q'}
       \bigl[R_A + R_E + R_S\bigr](\alpha J K Q K' Q')\,
       \rho^{K'}_{Q'}(\alpha J)\,.
  \end{aligned}
  \label{eq:SEE_ML}
\end{equation}

In Eqs.~\eqref{eq:SEE_MT} and \eqref{eq:SEE_ML}, $T_A,T_E,T_S$ are transfer rates and
$R_A,R_E,R_S$ are relaxation rates for absorption, spontaneous emission, and stimulated emission, respectively.
The multi-level rates are LL04 eqs.~7.14a--f, the multi-term rates are LL04
eqs.~7.41 and 7.45--7.46, and the
coherence-decay operator $N$ of Eq.~\eqref{eq:SEE_MT} is defined in LL04 eq.~7.39.

The absorption ($\eta^A_i$) and stimulated-emission ($\eta^S_i$) coefficients, and the emission vector
$\varepsilon_i=(2h\nu^3/c^2)\,\eta^S_i$ are defined by LL04 eq.~7.15 for the
multi-level description, and by LL04 eq.~7.47 for multi-term description.
The net propagation coefficients are $\eta_i=\eta^A_i-\eta^S_i$ and $\rho_i=\rho^A_i-\rho^S_i$.
The formal transfer step uses the constant-source and linear-source DELO forms
\citep{degl1985solution}.

In the self-consistent height-stratified radiation-field solve, the profile-averaged radiation tensor that enters
the SEE in the flat-spectrum limit is reconstructed at each depth from the polarized formal solution along a
discrete set of rays,
\begin{equation}
  \bar J^K_Q = \sum_{n} w_n\sum_{i=0}^{3}
    \mathcal T^K_Q(i,\hat{\bm\Omega}_n)\!\int\!\mathrm{d}\nu'\,\phi(\nu')\,
    I_i(\nu',\hat{\bm\Omega}_n)\,,
  \label{eq:app_reconstruction}
\end{equation}
This is the frequency-averaged discretization of Eq.~\eqref{eq:JKQ}; the profile integral over $\nu'$ removes the
explicit frequency dependence.
Here $\hat{\bm\Omega}_n$ and $w_n$ are the quadrature ray and weight, $\phi(\nu')$ is the normalized absorption
profile, and $I_i$ is the local Stokes vector at that depth along ray $n$.
The SEE is re-solved with this $J^K_Q$ and the loop iterated until $\max|\Delta\rho|$ falls below the tolerance.
An optional more advanced stopping criterion can also be used, which estimates the remaining fixed-point error as
$r/(1-\hat\lambda)$, where $r=\max|\Delta\rho|$ and $\hat\lambda$ is the estimated contraction rate inferred from
the recent residual decay.

\section{Benchmark details}
\label{sec:app_applications}

The atomic data follow the convention used by the selected atomic description.
For multi-term atoms, a radiative transition connects two terms and carries one multiplet Einstein coefficient.
For multi-level atoms, a radiative transition connects two fine-structure levels.
The He\,\textsc{i}~D$_3$ comparison in Fig.~\ref{fig:HeI_hazel} uses the helium atom model of
\textsc{Hazel2} \citep{asensio2008advanced}.
The model includes the triplet terms $2^3S$, $2^3P$, $3^3S$, $3^3P$, and $3^3D$.
The plotted line is the $3^3D_{1,2,3}\to 2^3P_{0,1,2}$ multiplet.
The constant-property slab setups match the \textsc{Hazel2} reference calculation.
The incident Allen continuum \citep{cox2015allen}, prescribed anisotropic radiation tensor, magnetic geometry, and
slab optical thickness are matched in each profile.

\begin{figure}
  \centering
  \includegraphics[width=0.5\linewidth]{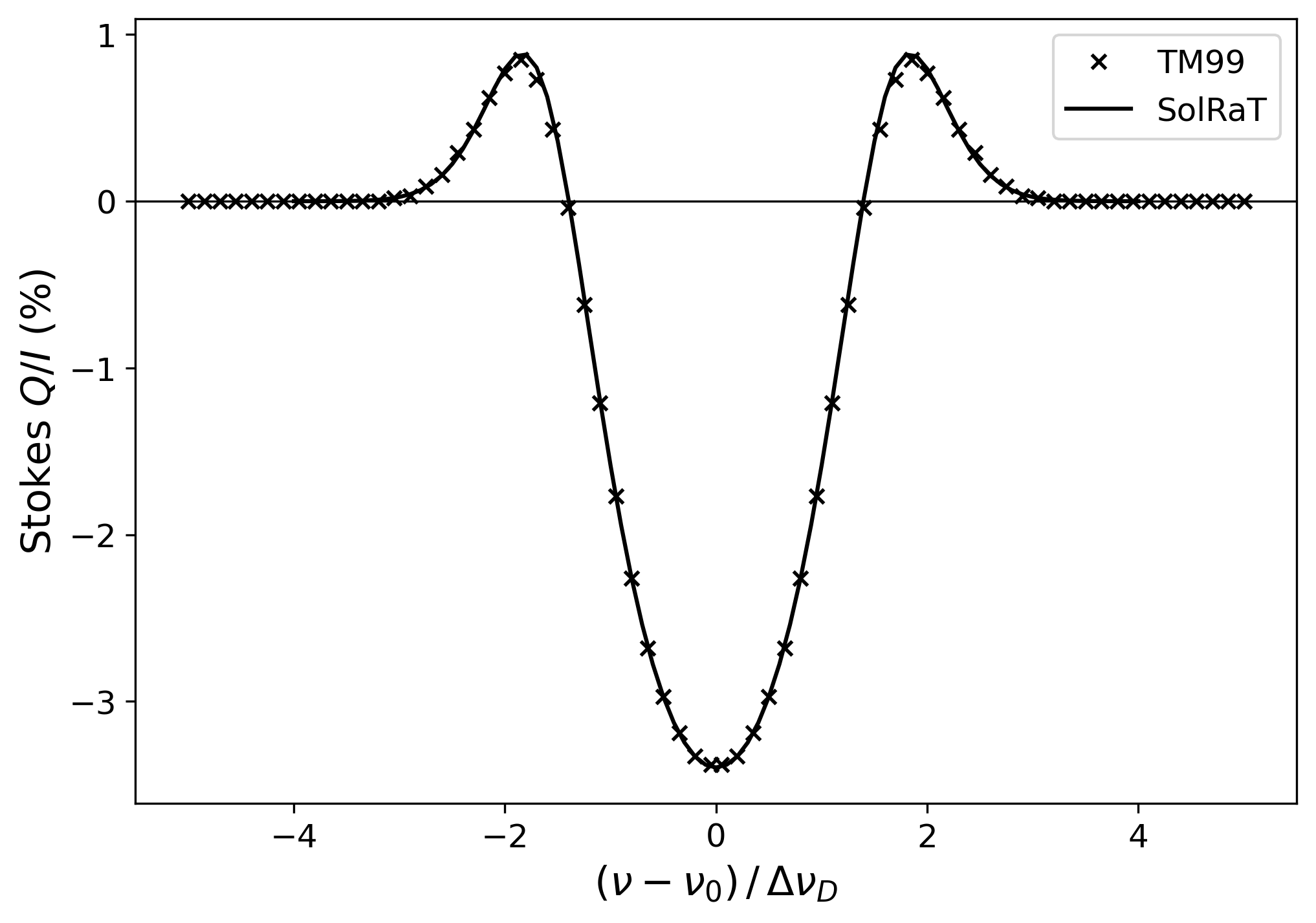}
  \caption{Emergent Stokes $Q/I$ in percent at $\mu=0.1$. Curve: \solrat.
    Crosses: digitized TM99 Fig.~10.}
  \label{fig:tm1999_qi}
\end{figure}

The TM99 resonance-scattering benchmark tests the self-consistent coupling between the SEE and the radiation
tensor in a $J=0\to1$ two-level atom in an isothermal slab.
The main text compares the converged alignment $\rho^2_0/\rho^0_0(\tau)$ for the two elastic depolarizing rates
used by TM99.
The associated emergent-profile comparison is presented in Fig.~\ref{fig:tm1999_qi},
which uses the TM99 Fig.~10 setup, with photon destruction probability $\epsilon=10^{-2}$,
elastic depolarizing rate $\delta^{(2)}=0$, and line-of-sight cosine
$\mu=0.1$.

Figures~\ref{fig:mt_ml_zeeman} and \ref{fig:mt_ml_scattering} use the same synthetic $^4P\to{}^4S$ multiplet.
The lower $^4S_{3/2}$ level is placed at $0$~cm$^{-1}$.
The upper $^4P_{5/2}$, $^4P_{3/2}$, and $^4P_{1/2}$ levels are placed at $20000.0$, $20001.0$,
and $20001.6$~cm$^{-1}$, respectively.
The plotted branch is $^4P_{5/2}\to{}^4S_{3/2}$, with the other two branches acting as nearby
fine-structure satellites.
The transition probability is $A_{ul}=10^7$~s$^{-1}$ and the adopted atomic mass is $56$~amu.
The two figures differ in field strength and illumination.
The LTE and non-LTE runs share the atomic model, the atmosphere, and the geometry.
The non-LTE curves in Fig.~\ref{fig:mt_ml_scattering} use a prescribed, cylindrically symmetric
Allen-continuum radiation tensor evaluated at $h=30''$ above the solar surface.
The radiation tensor is held fixed for this illustrative comparison.

\begin{figure}
  \centering
  \includegraphics[width=0.5\linewidth]{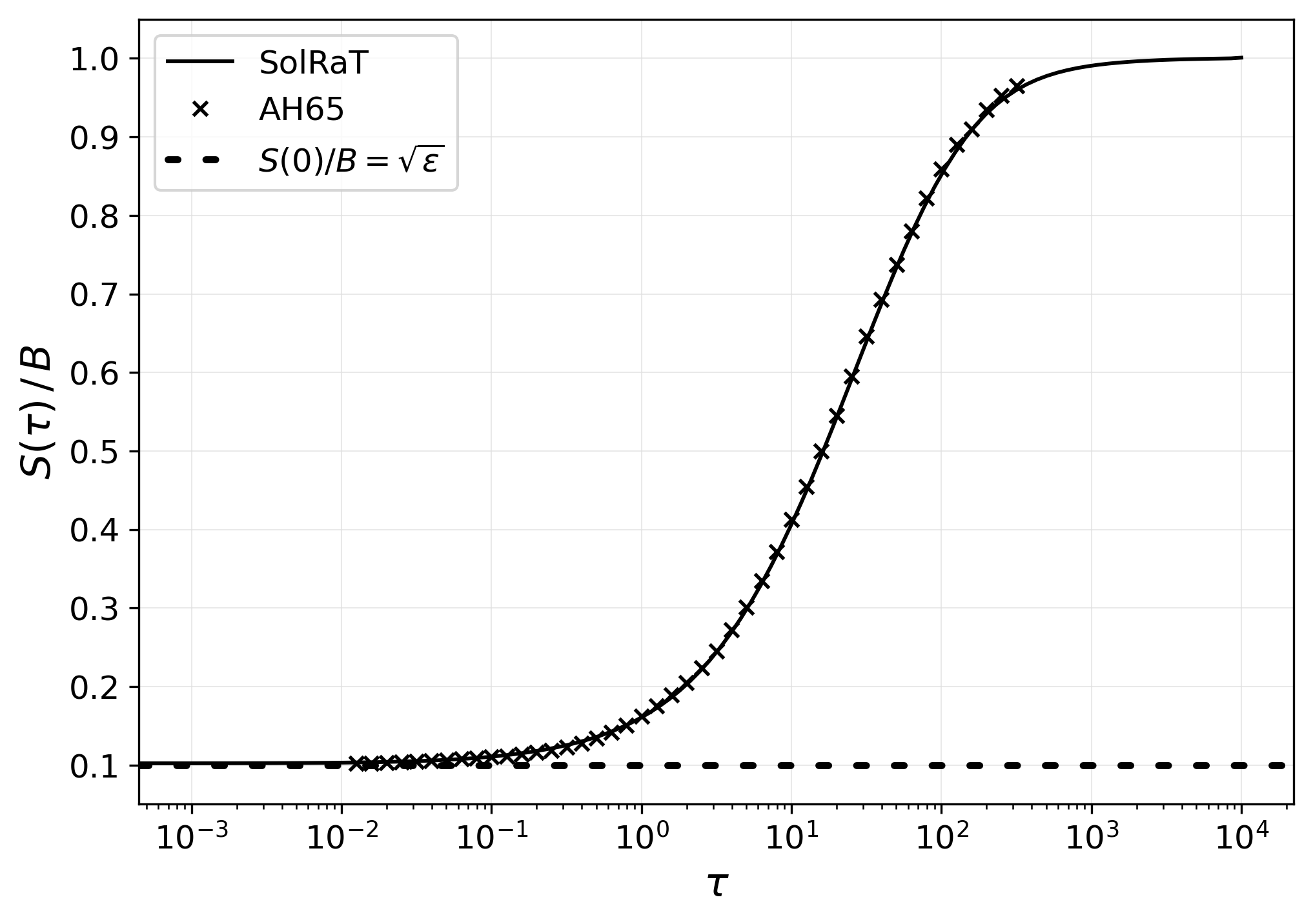}
  \caption{Line center source function for the AH65
    $\epsilon=10^{-2}$ benchmark. Solid line: \solrat. Crosses: digitized AH65
    Fig.~2. Dotted line: analytic asymptote $S(0)/B=\sqrt\epsilon$.}
  \label{fig:sqrt_epsilon}
\end{figure}

Figure~\ref{fig:sqrt_epsilon} shows the line center source function of a finite-$\epsilon$, isothermal,
semi-infinite two-level atom from the surface into the deep interior, using the pure-Doppler benchmark of AH65,
with the Planck function normalized to unity and infinite total optical thickness.
The agreement shows that the \solrat\ implementation reproduces this benchmark and confirms the expected
thermalization of a scattering transition with finite photon destruction probability.

\begin{figure}
  \centering
  \includegraphics[width=0.5\linewidth]{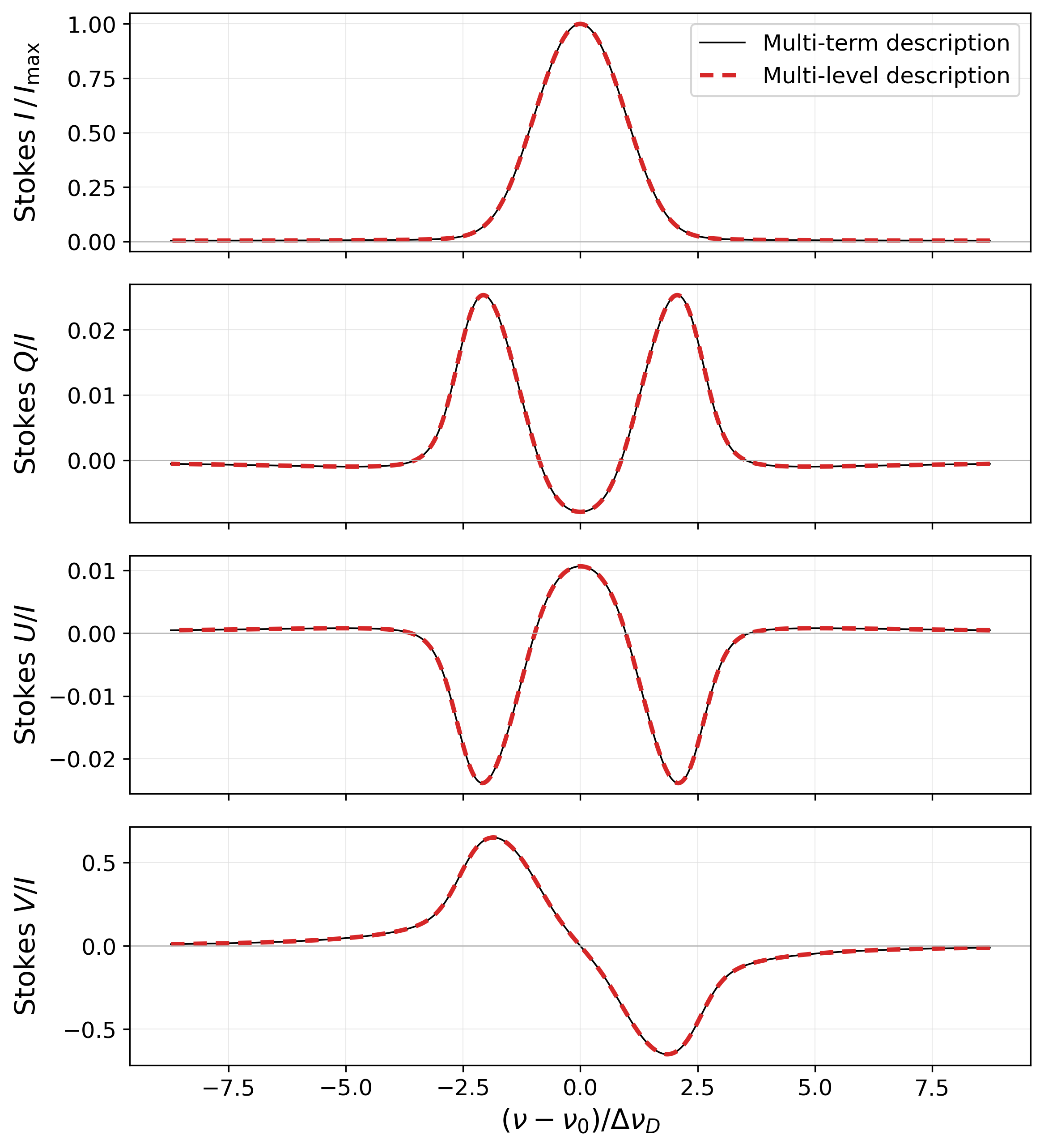}
  \caption{Stokes profiles for a transition between singlet terms.
    Solid lines: multi-term description. Dashed lines:
    multi-level description.}
  \label{fig:ml_vs_mt}
\end{figure}

For a transition between singlet terms ($S=0$), each term contains a single fine-structure level, so the multi-term
and multi-level descriptions are formally identical.
Figure~\ref{fig:ml_vs_mt} shows the resulting Stokes profiles for a prescribed-radiation synthesis in the same
atmosphere.
Their agreement to numerical precision validates this limiting case, and the same agreement is obtained inside the
self-consistent stratified solve.
Additional isolated checks compare the multi-level SEE with the analytic two-level resonance solution, the
multi-term RTE coefficients with a legacy implementation and with no-fine-structure analytic limits, and the
Wigner-symbol and Voigt routines with independent evaluations.

\begin{figure}
  \centering
  \includegraphics[width=0.5\linewidth]{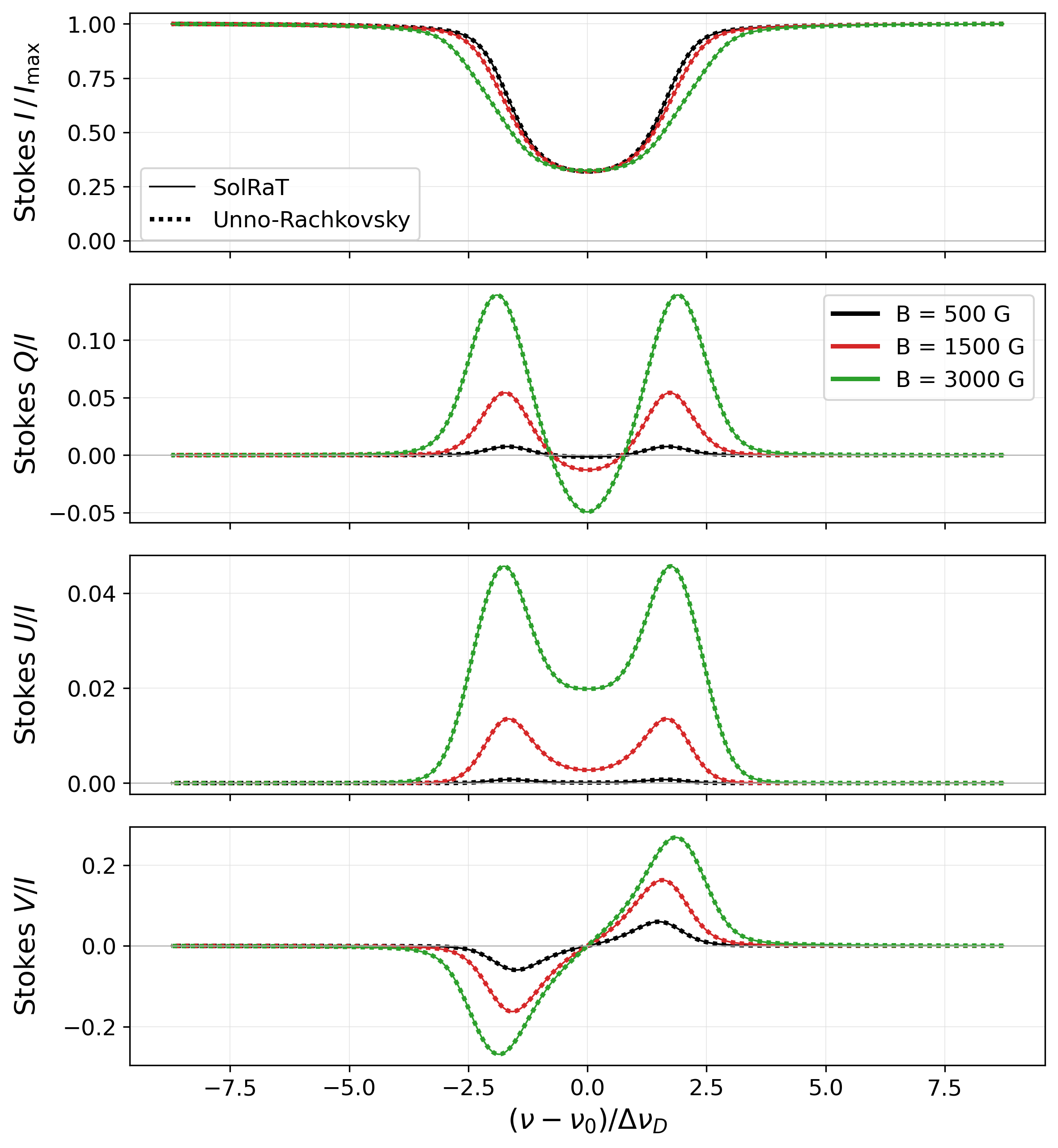}
  \caption{LTE Zeeman transfer for a $J=1\to0$ normal triplet. Solid lines:
    stratified prescribed-$J^K_Q$ \solrat. Dotted lines: Unno-Rachkovsky
    solution. Colors mark $B=500$, $1500$, and $3000$~G.}
  \label{fig:unno_rachkovsky}
\end{figure}

Figure~\ref{fig:unno_rachkovsky} compares the numerical prescribed-$J^K_Q$ stratified synthesis with the
classical LTE Milne-Eddington solution of \citet{unno1956line} and \citet{rachkovsky1962magnetic}.
The setup uses a linear source function and a depth-independent propagation matrix, for which the polarized solution
is known analytically.
The agreement tests the numerical RTE integration and the construction of the propagation matrix in the LTE Zeeman
limit.

\begin{figure}
  \centering
  \includegraphics[width=0.5\linewidth]{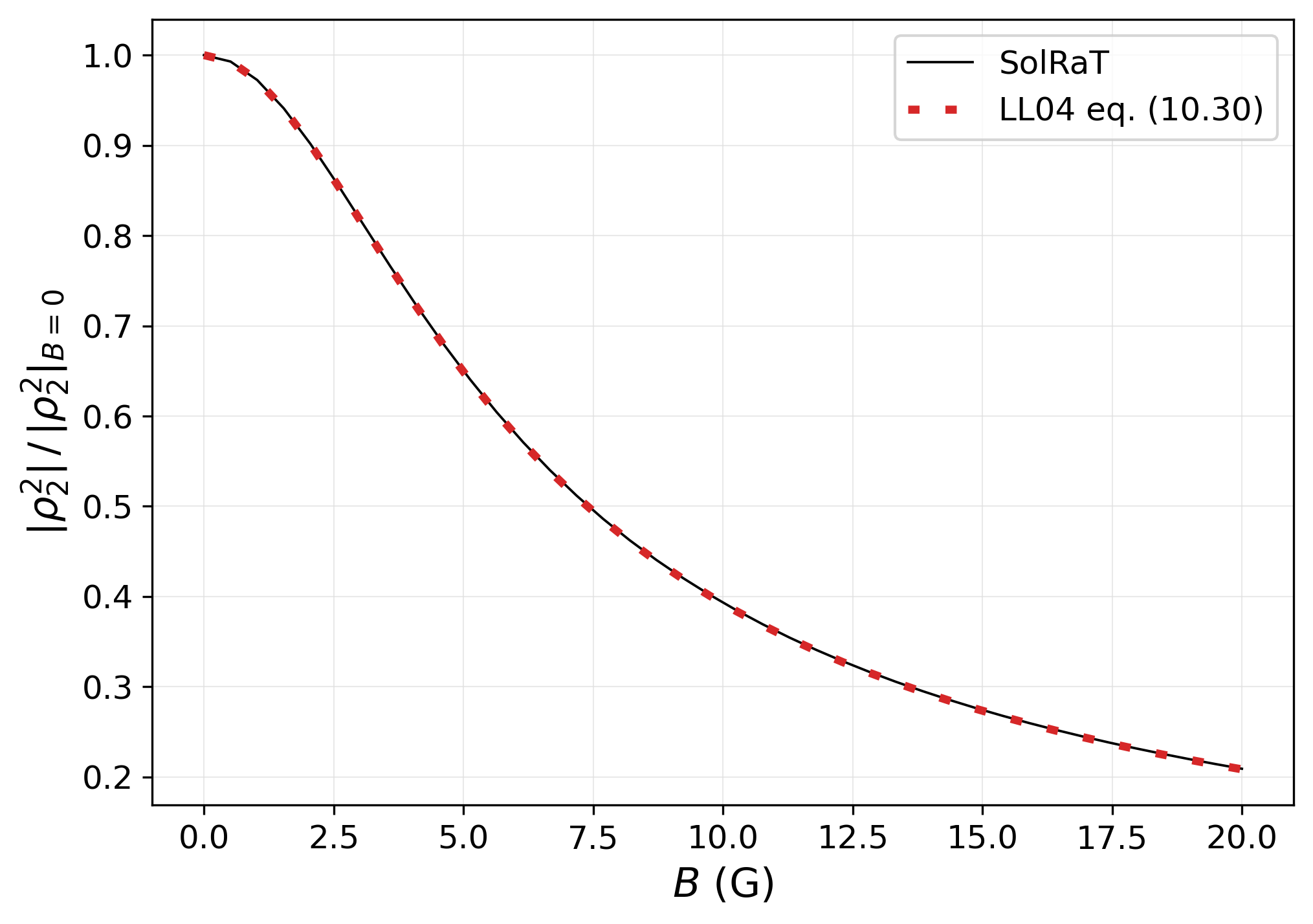}
  \caption{Hanle depolarization of the upper level coherence modulus for a
    resonance scattering transition. Solid line: \solrat. Dotted line:
    LL04 eq.~10.30.}
  \label{fig:validation}
\end{figure}

Figure~\ref{fig:validation} shows the Hanle depolarization of a resonance scattering transition.
With the geometry and incident radiation fixed, the only scanned quantity is the magnetic-field strength.
The normalized modulus of the upper-level coherence is therefore a direct check of the magnetic
coherence-relaxation operator against the LL04 analytic Hanle factor.

\section{Summation engine}
\label{sec:app_engine}

\solrat\ evaluates the constrained Wigner-symbol sums of the SEE and the RTE through a table representation of
the summation indices.
Each row corresponds to one admissible combination of angular-momentum indices, magnetic quantum numbers, and term
or transition labels.
Multiplicative factors are stored as columns, so triangular conditions, projection conservation, and vanishing
Wigner symbols act as filters on the table before the $3j$, $6j$, and $9j$ factors are evaluated.
The SEE and RTE are therefore implemented in an abstract form that resembles
the underlying mathematical expressions.
Changes to radiative or collisional rates, magnetic couplings,
or selected RTE components can therefore be reflected directly in the \solrat\ code,
while the required numerical optimization is applied by the engine.
This allows users to test alternative atomic descriptions, collision prescriptions, and transfer coefficients
without rewriting the error-prone optimized operations.

The mentioned table representation naturally supports staged caching.
All factors that depend only on the atomic structure are reduced once into fixed operators acting on the relevant
statistical-tensor components.
Self-consistent stratified calculations additionally apply an atmosphere-level cache for the fixed depth grid, angular quadrature,
frequency grid, and local thermodynamic and magnetic parameters.
During the self-consistent iteration, the statistical tensors and radiation tensor are updated while the atom-specific and
atmosphere-specific factors are reused.

\end{document}